\documentclass[a4paper,UKenglish,cleveref, autoref, thm-restate]{lipics-v2021}

\title{PoTo: A Hybrid Andersen's Points-to Analysis for Python}

\author{Ingkarat Rak-amnouykit}{Rensselaer Polytechnic Institute, USA}{rakami@rpi.edu}{https://orcid.org/0000-0001-7087-9282}{}
\author{Ana Milanova}{Rensselaer Polytechnic Institute, USA}{milanova@cs.rpi.edu}{https://orcid.org/0009-0005-9667-5276}{}
\author{Guillaume Baudart}{DI ENS, Ecole normale supérieure, PSL University, CNRS, INRIA, France}{guillaume.baudart@inria.fr}{https://orcid.org/0000-0003-2230-1616}{}
\author{Martin Hirzel}{IBM Research, USA}{hirzel@us.ibm.com}{https://orcid.org/0009-0006-8840-6065}{}
\author{Julian Dolby}{IBM Research, USA}{dolby@us.ibm.com}{https://orcid.org/0000-0002-6658-2678}{}

\authorrunning{I. Rak-amnouykit, A. Milanova, G. Baudart, M. Hirzel, and J. Dolby} 

\Copyright{Ingkarat Rak-amnouykit, Ana Milanova, Guillaume Baudart, Martin Hirzel, and Julian Dolby} 


\category{} 

\relatedversion{} 

\acknowledgements{We thank the ECOOP reviewers for their valuable comments and the US National Science Foundation for supporting our work through awards \#1814898 and \#2232061.}

\nolinenumbers 

\EventEditors{Jonathan Aldrich and Alexandra Silva}
\EventNoEds{2}
\EventLongTitle{39th European Conference on Object-Oriented Programming (ECOOP 2025)}
\EventShortTitle{ECOOP 2025}
\EventAcronym{ECOOP}
\EventYear{2025}
\EventDate{June 30--July 2, 2025}
\EventLocation{Bergen, Norway}
\EventLogo{}
\SeriesVolume{333}
\ArticleNo{26}

\usepackage{algorithm2e}
\usepackage{algorithmic}
\usepackage{booktabs}
\usepackage{tikz}
\usetikzlibrary{calc}

\usepackage{xcolor}
\definecolor{keyword}{HTML}{180FE1}
\definecolor{operator}{HTML}{A51DFF}
\definecolor{string}{HTML}{C03333}
\definecolor{background}{HTML}{F7F7F7}

\begin{document}

\definecolor{keyword}{HTML}{180FE1}
\definecolor{operator}{HTML}{A51DFF}
\definecolor{string}{HTML}{C03333}
\definecolor{background}{HTML}{F7F7F7}

\newcommand{\pit}[1]{\ensuremath{\textit{#1}}}
\newcommand{\ptt}[1]{\ensuremath{\textsf{#1}}}
\newcommand{\pkw}[1]{\ensuremath{\textcolor{keyword}{\ptt{#1}}}}


\newcommand{\pyimport}[1]{\ensuremath{\pkw{import}\ #1}}
\newcommand{\pyfrom}[2]{\ensuremath{\pkw{from} \ #1\ \pkw{import}\ #2}}
\newcommand{\pyas}[2]{\ensuremath{#1 \ \pkw{as} \ #2}}
\newcommand{\pyfun}[3]{\ensuremath{\pkw{def}\ #1 \ptt{(}#2\ptt{):}\ #3}}
\newcommand{\pyclass}[3]{\ensuremath{\pkw{class}\ #1 \ptt{(}#2\ptt{):}\ #3}}
\newcommand{\pymodule}[2]{\ensuremath{\pkw{module}\ \ptt{#1}:\ #2}}

\newcommand{\pypass}{\ensuremath{\pkw{pass}}}
\newcommand{\pyseq}[2]{\ensuremath{#1\ \ptt{;}\ #2}}
\newcommand{\pyraise}[1]{\ensuremath{\pkw{raise}\ #1}}
\newcommand{\pyassign}[2]{\ensuremath{#1\ \ptt{=}\ #2}}
\newcommand{\pyite}[3]{\ensuremath{\pkw{if}\ #1\ \ptt{:}\ #2 \ \pkw{else}\ptt{:}\ #3}}
\newcommand{\pyfor}[3]{\ensuremath{\pkw{for}\ #1\ \ptt{in}\ #2 \ptt{:}\ #3}}
\newcommand{\pyreturn}[1]{\ensuremath{\pkw{return}\ #1}}
\newcommand{\pyother}[1]{\ensuremath{\pit{Other}\ (#1)}}

\newcommand{\pynone}{\ensuremath{\ptt{None}}}
\newcommand{\pycall}[2]{\ensuremath{#1\ptt{(}#2\ptt{)}}}
\newcommand{\pyattr}[2]{\ensuremath{#1\ptt{.}#2}}
\newcommand{\pysub}[2]{\ensuremath{#1\ptt{[}#2\ptt{]}}}
\newcommand{\pyget}[3]{\ensuremath{#1\ptt{.get(}#2\ptt{,} \ #3\ptt{)}}}
\newcommand{\pytuple}[1]{\ensuremath{\ptt{(}#1\ptt{)}}}
\newcommand{\pylist}[1]{\ensuremath{\ptt{[}#1\ptt{]}}}
\newcommand{\pydict}[1]{\ensuremath{\ptt{\{}#1\ptt{\}}}}
\newcommand{\pykeyval}[2]{\ensuremath{#1\ptt{:}\ #2}}
\newcommand{\pyunop}[1]{\ensuremath{\pit{Unop}\ptt{(}#1\ptt{)}}}
\newcommand{\pybinop}[2]{\ensuremath{#1\ \pit{op}\ #2}}
\newcommand{\pycompare}[2]{\ensuremath{#1\ \pit{cop}\ #2}}
\newcommand{\pystar}[1]{\ensuremath{\ptt{*}#1}}
\newcommand{\pystarstar}[1]{\ensuremath{\ptt{**}#1}}

\newcommand{\pylistcomp}[4]{\ptt{[}\ensuremath{#1\ \pkw{for}\ #2\ \pkw{in}\ #3\ \pkw{if}\ #4} \ptt{]}}
\newcommand{\pydictcomp}[5]{\ptt{\{}\ensuremath{#1\ptt{:}#2\ \pkw{for}\ #3\ \pkw{in}\ #4\ \pkw{if}\ #5} \ptt{\}}}

\newenvironment*{stack}{\begin{array}[t]{@{}l@{}}}{\end{array}}
\newcommand{\iComp}[2]{\mathcal{I}(#1, #2)}
\newcommand{\iRet}[2]{\pkw{return} \ \left(#1, #2\right)}
\newcommand{\set}[1]{\left\{ #1 \right\}}
\newcommand{\env}[1]{\left[ #1 \right]}

\newcommand{\algComment}[1]{\textcolor{black!50!white}{\# \textit{#1}}}

\newcommand{\python}[1]{\lstinline[language=python, basicstyle=\small\ttfamily]{#1}}

\newcommand{\todo}[1]{{\color{red}\bfseries [#1]}}
\newcommand{\todobill}[1]{{\color{blue}\bfseries [#1]}}
\newcommand{\ana}[1]{\todo{Ana: #1}}
\newcommand{\bill}[1]{\todo{Bill: #1}}
\newcommand{\dan}[1]{\todo{Dan: #1}}
\newcommand{\julian}[1]{\todo{Julian: #1}}
\newcommand{\martin}[1]{\todo{Martin: #1}}    
\newcommand{\guillaume}[1]{\todo{Guillaume: #1}} 

\newcommand{\code}[1]{{\sf #1}\xspace}

\newcommand*{\errcode}[1]{\textsc{#1}}

\lstnewenvironment{pythonn}{}{}
\newcommand*{\lbr}{\texttt{\{}}
\newcommand*{\rbr}{\texttt{\}}}
\newcommand*{\pyplain}[1]{\texttt{\small #1}}
\newcommand*{\pystring}[1]{\texttt{\small\textcolor{string}{#1}}}

\maketitle

\begin{abstract}
As Python is increasingly being adopted for large and complex programs, the importance of static analysis for Python (such as type inference) grows.
Unfortunately, static analysis for Python remains a challenging task due to its dynamic language features and its abundant external libraries.
To help fill this gap, this paper presents PoTo, an Andersen-style context-insensitive and flow-insensitive points-to analysis for Python. 
PoTo addresses Python-specific challenges and works for large programs via a novel hybrid evaluation, integrating traditional static points-to analysis with concrete evaluation in the Python interpreter for external library calls.
We evaluate the analysis with two clients: type inference and call graph construction.
This paper presents PoTo+, a static type inference for Python built on PoTo.
We evaluate PoTo+ and compare it to two state-of-the-art Python type inference techniques: (1)~the static rule-based Pytype and (2)~the deep-learning based DLInfer. Our results show that PoTo+ outperforms both Pytype and DLInfer on existing Python packages.
This paper also presents PoToCG, a call-graph construction analysis for Python built on PoTo.
We compare PoToCG to PyCG, the state of the art for this problem, and show that PoTo
produces more complete and more precise call graphs.
\end{abstract}


\section{Introduction}
\label{sec:introduction}

Points-to analysis is a fundamental static analysis that determines what objects 
a reference variable may point to. It has a wide variety of applications (aka.\ \emph{client} analyses), including call graph
construction, type inference, debugging, and security vulnerability detection. Essentially all
interesting questions one could ask about a program require some form of points-to
information and points-to analysis has been studied extensively in the context of 
mainstream languages such as C, C++, Java, and JavaScript. 
Surprisingly, points-to analysis for Python has received less attention, 
despite Python's popularity. 

This paper presents PoTo, an Andersen-like point-to analysis for Python. Andersen's 
analysis~\cite{andersen_1994} is a classical flow-insensitive, context-insensitive, and inclusion-constraint-based analysis.
It computes a points-to graph $\mathit{Pt}$ where the nodes are reference variables and heap objects, 
and the edges represent points-to relations. For example, an edge $\pit{x} \rightarrow o$ 
represents that \pit{x} points to heap object $o$ (i.e., reference \pit{x} stores the address of
$o$), and \raisebox{-0.5mm}{$o_1 \stackrel{f}{\rightarrow} o_2$} represents that field $f$ of $o_1$ points to $o_2$.

While PoTo leverages well-known techniques, it also adapts to the unique 
challenges of Python: complex syntax and module system, dynamic semantics, 
and ubiquitous use of external libraries whose code is unavailable. 
First, our solution presents principled translation from Python source into the 3-address code 
intermediate representation demanded by points-to analysis. It specifies explicitly 
which Python features are handled precisely and faithfully to the feature semantics 
and which are handled approximately via a default interpretation; due to the complexity of Python,
handling is unsound. 
Second, a novelty of PoTo is \emph{hybridization}. We observe that many expressions, particularly ones 
stemming from external libraries and built-in functions, are available for \emph{concrete evaluation} 
but are unavailable for standard \emph{abstract evaluation} by virtue of their code being unavailable. 
Therefore we \emph{concretely evaluate} the expressions and propagate the concrete objects 
through the points-to graph. This yields more information during inclusion constraint resolution 
and improves program coverage.

We evaluate PoTo on concrete type inference and call graph construction. We choose type inference because there has been 
significant interest in the problem in recent years. In addition to traditional approaches such as Pytype~\cite{pytype_2016}, 
there has been significant advance in deep-learning-based approaches, e.g.,~\cite{allamanis_et_al_2020,pradel_et_al_2020,yan_et_al_2023}. 
PoTo+ is the type inference client that largely draws types from PoTo's result: given $\mathit{Pt}(\pit{x})$ computed 
by the points-to analysis, the type of \pit{x} is the union of the types of objects $o \in \mathit{Pt}(\pit{x})$. 
We compare PoTo+ against (1)~Pytype~\cite{pytype_2016}, arguably the most advanced rule-based type 
inference tool, and (2)~DLInfer~\cite{yan_et_al_2023}, a state-of-the-art deep-learning-based tool, on a benchmark suite of 
Python packages ranging from 3,556 to 285,515 lines of code. 
Our results show that PoTo+ is comparable to Pytype and both techniques outperform DLInfer in terms of coverage (i.e., percentage of variables for which a type is reported) and 
correctness of inferred types. Furthermore, PoTo scales better than Pytype. Call graph construction is a classical client of points-to analysis.
We build PoToCG based on PoTo and compare it to PyCG~\cite{salis_et_al_2021}, the leading call graph construction analysis for Python.
The results show that PoToCG
constructs more complete and more precise call graphs compared to PyCG and it scales better.

\medskip

The contributions of our work are as follows:
\begin{itemize}
\item PoTo, the first Andersen-style points-to analysis for Python.
\item Hybridization weaving concrete and abstract evaluation.
\item Evaluation on two clients of points-to analysis: type inference and call graph construction.
\end{itemize}

Ultimately, we hope that this work contributes to tools that make
Python developers more productive and Python code more robust.

\section{Problem Statement and Overview}
\label{sec:overview}


The problem at hand is to design an Andersen-style points-to analysis for Python and \emph{scale the analysis to large real-world Python packages}. Andersen's points-to analysis is a classical static analysis problem. 
It is a whole-program flow-insensitive and context-insensitive analysis that tracks flow of values via inclusion constraints. E.g., an assignment statement \python{x = y} triggers an inclusion constraint $\mathit{Pt}(\ptt{y}) \subseteq \mathit{Pt}(\ptt{x})$ indicating that the points-to set of \python{y} flows to the points-to set of \python{x}. 

There are numerous challenges for Andersen's analysis and more generally static analysis for Python. First, Andersen's analysis is defined over a \mbox{3-address} code intermediate representation (IR) serving as the foundation for the inclusion constraints. Translation from high-level C, C++, Java, and JavaScript code into \mbox{3-address} code is well-studied and there are mature tools that provide the translation, notably LLVM~\cite{lattner_adve_2004}, Soot~\cite{valleerai_2000}, WALA~\cite{wala_2006}, and Doop~\cite{smaragdakis_bravenboer_2011}. Surprisingly, static analysis for Python is largely ad-hoc AST-based analysis with each new work (e.g., \cite{pytype_2016,yan_et_al_2023}) embedding its own interpretation of Python AST constructs and translation (if any) of those constructs into a \mbox{3-address} code IR. Translating Python AST constructs into a \mbox{3-address} code IR is challenging due to (1)~Python's complex syntax and dynamic semantics and (2)~Python's rich module system and scoping issues arising from it.  
Python allows for reflection and metaprogramming, subclassing conditioned on runtime values, complex With statements and context managers, and many other features~\cite{yang_milanova_hirzel_2022} that are difficult or impossible to handle soundly in a static semantics. Even a construct such as Subscript, e.g., \mbox{\python{a[index_expr]}}, is non-trivial in the sense that its canonical interpretation as element access of a list-like structure is unsound (we choose the canonical interpretation for both Subscript and Attribute; a more faithful but also more complex interpretation would be to treat \python{__getitem__} and \python{__getattr__} as virtual calls). A principled analysis therefore must choose a subset of constructs to interpret precisely. We present a minimal syntax that specifies a subset of Python, and for that subset we present a precise interpretation that is faithful to the Python construct's semantics. All other Python AST constructs are abstracted by a fall-through structure \textit{Other} and receive a default interpretation that can be both over-approximate and unsound. Going forward we sometimes refer to precisely-interpreted constructs as \emph{interpreted} and to \textit{Other} as \emph{uninterpreted}. A key goal is to describe the points-to analysis as completely as possible, grounding the syntax and specifying interpretation semantics for both kinds of constructs, while at the same time abstracting away unnecessary detail. ~\Cref{sec:syntax_and_semantics} specifies interpreted and uninterpreted constructs and their semantics, and makes it clear that we support memory-accesses and nested functions. It also explicitly lists the features under \pit{Other}, e.g., generator expressions and \textsf{\small With} statements.

Another challenge is use of external libraries. Whole-program static analysis generally assumes that library code is available. However, Python has a particularly extensive eco-system of open-source libraries; it is a dynamically-typed language where library interfaces usually lack type annotations; Python libraries and their interfaces are often rapidly evolving; and libraries are often at least partially implemented in C instead of Python. We address this with a novel hybridization technique that weaves concrete and abstract evaluation during 3-address code generation and constraint resolution.  

While to the best of our knowledge we are the first to present a points-to analysis for Python, the problem has been studied extensively in mainstream languages, notably C, C++, Java, and JavaScript. (We detail related work in~\Cref{sec:related}.) JavaScript is a dynamically typed language, as is Python. However, JavaScript poses unique challenges and program analysis aims to address these challenges: event-driven nature of applications, interaction with the DOM, runtime construction of object property strings, and runtime creation and deletion of object properties. None of these is a significant issue for Python, which increasingly is a general-purpose language used to build libraries and applications similarly to Java. We therefore model Andersen's analysis after Java's analysis, extracting and analyzing a Java-like subset of Python. Python presents new challenges though, including lack of a 3-address code IR, missing external libraries, and first-class functions and classes. 

\begin{figure}
\centering
\noindent
\begin{subfigure}[t]{0.42\textwidth}
\begin{lstlisting}
import re

def url_regex():
  regex_cache = None
  if regex_cache is None:
    regex_cache = re.compile(r"p")
  return regex_cache

def str_validator(value):
  if isinstance(value,str):
    return value.value
  else:
    return value
    
def validate(value):
  url = str_validator(value)
  m = url_regex().match(url)
  if m.end() != len(url): 
    raise Exception(m.end())

def main():
  validate("p abcd")
 
\end{lstlisting}
\vspace{1.15em}
\caption{Python source}
\end{subfigure}
\hfill
\begin{subfigure}[t]{0.57\textwidth}
\begin{lstlisting}[breaklines=false]
import re # External import

def url_regex():
  regex_cache = None
  regex_cache=(const,<class 're.Pattern'>,...)
  url_regex_ret = regex_cache

def str_validator(value):
  str_validator_ret = value.value
  str_validator_ret = value

def validate(value):
  url = str_validator(value)
  t1 = url_regex()
  t2 = t1.match  
  m = t2(url) 
  t3 = m.end 
  t4 = t3() 
  t5 = len(url)
  t6 = (const,<class 'Exception'>,...) 
  t7 = m.end 
  t8 = t7() 
  t9 = t8(t7)

\end{lstlisting}
\caption{3-address code for (a), excluding \python{main}}
\end{subfigure}
\vspace*{-2mm}
\caption{Illustrating example, adapted from DLInfer~\cite{yan_et_al_2023}.}
\label{fig:overview-example}
\end{figure}

Consider the example in Figure~\ref{fig:overview-example}, adapted from DLInfer, a recent paper on neural type inference for Python.
Our analysis has two core phases~(PoTo), followed by a client phase:

\begin{description}
  \item[Phase 1:] Python source $\longrightarrow$ 3-address code
  \item[Phase 2:] 3-address code $\longrightarrow$ Points-to graph
  \item[Phase 3 (client):] Points-to graph $\longrightarrow$ Concrete type inference or call graph
\end{description}

Phase~1 takes as input Python source code and produces \mbox{3-address} code; this phase works at the granularity of a function.
Phase~2 processes the \mbox{3-address} statements as inclusion constraints computing the points-to graph.
The two phases are intertwined --- roughly, the analysis starts at the main function and immediately invokes Phase~1 on main generating \mbox{3-address} code for main; it then invokes Phase~2 to solve the \mbox{3-address} code for main.
As new functions become reachable, the analysis invokes \mbox{3-address} code generation, then it proceeds to solve the constraints until the points-to graph reaches a fixpoint.
As expected, one can implement many client analyses on top of the points-to results. 
Phase~3 focuses on concrete (i.e., non-polymorphic) type inference and call graphs as client analyses. 

\begin{figure}
\centerline{\begin{minipage}{0.5\textwidth}
\[
\begin{array}{lll}
1. & \ptt{x = object} & \mathit{new~assignment} \\
2. & \ptt{x = y} & \mathit{copy~propagation} \\
3. & \ptt{x.f = y} & \mathit{field~write} \\
4. & \ptt{x = y.f} & \mathit{field~read} \\
5. & \ptt{x = y(z)} & \mathit{closure~call} 
\end{array}
\]
\end{minipage}}
\caption{3-address code statements.}
\label{fig:3-address-code}
\end{figure}

The first phase does principled translation of Python AST constructs into the standard \mbox{3-address} statements shown in Figure~\ref{fig:3-address-code}.
For example, the Python source in Figure~\ref{fig:overview-example}(a) Line~17 translates into the \mbox{3-address} sequence of statements in Figure~\ref{fig:overview-example}(b) Lines 14--16.
A novelty in our treatment is the weaving of concrete evaluation into \mbox{3-address} code translation and later constraint resolution.
During translation, the analysis concretely evaluates every expression in its enclosing import environment.
If evaluation succeeds, translation returns the resulting constant; otherwise, it proceeds recursively and returns the corresponding \mbox{3-address} code statements.
For example, in Figure~\ref{fig:overview-example}(a) Line~6, the right-hand-side evaluates into a concrete object: \python{(const,<class 're.Pattern'>,...)} in Figure~\ref{fig:overview-example}(b) Line~5 (we write a concrete object as a triple of \python{const}, the object type, and its value, with the value typically elided). 
As is standard for flow-insensitive analysis, the translation ignores control flow and basic blocks.
Each function in Figure~\ref{fig:overview-example}(b) is a straight-line sequence of \mbox{3-address} statements.
Suffix \python{_ret} indicates the function's return value (e.g., \python{str_validator_ret}).



The second phase of the analysis solves the \mbox{3-address} statements as inclusion constraints, incorporating (1) class and function objects as first-class values, and (2) concrete evaluation. The analysis maintains \emph{abstract objects} (constructed from code in the package under analysis) and \emph{concrete objects} (constructed from concrete evaluation). Abstract objects, in turn, fall into three categories: \emph{meta-class} objects, \emph{meta-func} objects (more precisely closure objects), and \emph{data} objects. The analysis maintains an abstract reference environment, which is the package-under-analysis environment, and uses it to resolve names defined in the package. 

When processing a call or a field access statement, the analysis retrieves each object in the points-to set of the receiver variable. Roughly speaking, if the function object at the call is an abstract one, the analysis proceeds with abstract evaluation. If it is a concrete one, it attempts concrete evaluation. For example, in \mbox{\python{url = str_validator(value)}}, the analysis examines the points-to set of \python{str_validator} and retrieves the abstract function 
object representing the \python{str_validator} function. This triggers abstract evaluation and~(standard) addition of the points-to set of \python{str_validator_ret} (the special return variable) to the points-to set of the left-hand-side variable \python{url}. Eventually, the points-to set of \python{url} becomes \mbox{\{\python{(const, <class 'str'>, 'p abcd')}\}}. Concrete evaluation of \python{t2 = t1.match} returns the concrete closure object corresponding to the match function, and finally concrete evaluation of \mbox{\python{m = t2(url)}} returns a concrete \python{re.Match} object. The hybrid analysis infers concrete types for identifiers \python{url} and \python{m}, respectively \python{<class 'str'>} and \python{<class 're.Match'>}, while Pytype reports only the fall-through type Any for both.




\section{Syntax and Semantics}
\label{sec:syntax_and_semantics}

Recent work defines syntax for a subset of Python and a corresponding interpretation semantics for the purpose of weakest precondition inference~\cite{rakamnouykit_et_al_2024}. A notable idea in this work is the separation of Python constructs into interpreted and uninterpreted ones. 
In this paper, we follow the idea of separating constructs into interpreted and $\textit{Other}$ but differ in important ways as our goal is interprocedural flow- and context-insensitive points-to analysis rather than weakest precondition inference. Unlike~\cite{rakamnouykit_et_al_2024}, we track object creation and flow of values and construct a call graph on the fly. Flow-insensitive points-to analysis demands a different set of interpreted constructs and a different interpretation.


Section~\ref{sec:syntax} defines the syntax for the purposes of flow-insensitive points-to analysis, 
Section~\ref{sec:3-address_code} defines the interpretation that generates 3-address code, 
and Section~\ref{sec:constraint_resolution} presents Andersen-style constraint resolution, highlighting Python-specific semantics.


\subsection{A Minimal Python Syntax}
\label{sec:syntax}

\begin{figure}
$$
\begin{array}{l@{}r@{\ }ll}

e &::=& c
\mid  x
\mid \pyattr{e}{x}
\mid \pysub{e}{e}
\mid   \pycall{e}{e, ..., e} & \ptt{const} \mid \ptt{Name} \mid \ptt{Attribute} \mid \ptt{Subscript} \mid \ptt{Call} \\ 
&\mid& \pylist{e, ..., e}
\mid \pydict{\pykeyval{e}{e}, ..., \pykeyval{e}{e}} 
\mid \pytuple{e, ..., e} & \ptt{List} \mid \ptt{Dictionary} \mid \ptt{Tuple} \\
&\mid& \pylistcomp{e}{x, ..., x}{e}{e}  &\ptt{ListComp} \\
&\mid& \pydictcomp{e}{e}{x, ..., x}{e}{e} & \ptt{DictComp}  \\
&\mid& \pybinop{e}{e} \mid \pycompare{e}{e} & \ptt{BinOp} \mid \ptt{Compare} \\
&\mid& \pyother{e, ..., e} & \pit{All other Python AST expressions} \\
\\

s &::=& 
\pypass
\mid   \pyassign{x}{e}
\mid   \pyassign{\pyattr{e}{x}}{e}
\mid   \pyassign{\pysub{e}{e}}{e} & \ptt{Pass} \mid \ptt{Assign} \\
&\mid&  \pyseq{s}{s} \mid   \pyfor{e}{e}{s} & \ptt{Suite} \mid \ptt{For} \\
&\mid& \pyfun{f}{x, ..., x}{\pyseq{s}{\pyreturn{e}}} & \ptt{FunctionDef} \mid \ptt{Return}\\
&\mid& \pyclass{C}{e, ..., e}{s} & \ptt{ClassDef} \\

&\mid& \pyother{s, ..., s} & \pit{All other Python AST statements} \\
\\
i &::=& \pyimport{p}\ (\pkw{as}\ x)? 
 \mid \pyfrom{p}{x} \ (\pkw{as}\ x)? & \ptt{Import} \mid \ptt{ImportFrom} \\
 &\mid& \pyseq{i}{i}  \\
\\
m &::=& \pyseq{i}{s} & \ptt{Module}
\end{array}
$$
\caption{Syntax of a subset of Python. The non-terminal names on the right are based on the official Python abstract syntax tree (AST); e.g., the AST node for a statement sequence is \textsf{\small Suite}.} 
\label{Fig:syntax_petpy}
\end{figure}


\Cref{Fig:syntax_petpy} specifies the syntax which grounds the analysis. 
An expression $e$ can be a constant~$c$ (\python{42} or \python{"foo"}), a variable (\python{x}), an attribute access (\python{x.foo}), a subscript access (\python{x["bar"]}), a list, a tuple, a dictionary, a list comprehension (\python{[2*x for x in range(10)]}), a dictionary comprehension (\python{\{x:f(x) for x in range(x)\}}), a binary operation or a compare operation.
All other Python expressions are under $\pyother{e_1, ..., e_n}$ and receive the default interpretation; their AST nodes are \textsf{\small BoolOp}, \textsf{\small NamedExpr}, \textsf{\small UnaryOp}, \textsf{\small Lambda}, \textsf{\small IfExp}, \textsf{\small Set}, \textsf{\small SetComp}, \textsf{\small GeneratorExp}, \textsf{\small Await}, \textsf{\small Yield}, \textsf{\small YieldFrom}, \textsf{\small FormattedValue}, \textsf{\small JoinedStr}, \textsf{\small Starred}, and \textsf{\small Slice}.

A statement $s$ can be $\pypass$, the assignment of either a variable (\python{x = 42}), an attribute (\python{x.foo = 42}), or a subscript (\python{x["bar"] = 42}), a sequence, a loop, a function definition, or a class definition.
All other Python statements are captured under $\pyother{s_1, ..., s_n}$ and receive the default interpretation; their AST nodes are \textsf{\small Delete}, \textsf{\small TypeAlias}, \textsf{\small AugAssign}, \textsf{\small AnnAssign}, \textsf{\small AsyncFor}, \textsf{\small While}, \textsf{\small If}, \textsf{\small With}, \textsf{\small AsyncWith}, \textsf{\small Match}, \textsf{\small Raise}, \textsf{\small Try}, \textsf{\small TryStar}, \textsf{\small Assert}, \textsf{\small Global}, \textsf{\small Nonlocal}, \textsf{\small Break}, and \textsf{\small Continue}.
This default interpretation is often all that is needed. For example,
\textsf{\small Break} and \textsf{\small Continue} are irrelevant to
flow-insensitive analysis.
As another example, for \textsf{\small If} statements, the default
interpretation descends into the AST node of the children and processes them.
As illustrated in \Cref{fig:overview-example}, this is adequate for
flow-insensitive analysis.
To simplify the presentation, 
we leave out the syntax of complex left-hand-side expressions in assignments (e.g., \python{a, *(b, c) = (1, (2, 3), 4)}), variable-length arguments, and keywords arguments. We use deconstruct-and-bind preprocessing to reduce these into sequences of assignments as in~\Cref{Fig:syntax_petpy}. For example, \python{a, *(b, c) = (1, (2, 3), 4)} reduces into a sequence including \python{a = 1}, \python{b = (2, 3)} and \python{c = 4}. 


A module $m$ starts with a sequence of import $i$, $\pyimport{p}$ or $\pyfrom{p}{x}$, with an optional alias name $(\pkw{as}\ x)$, followed by a statement.
A package under analysis is a sequence of modules.

\subsection{3-address Code Generation}
\label{sec:3-address_code}



3-address-code translation takes as input the Python source (i.e., AST) of a function and produces a sequence of 3-address-code statements of the form in Figure~\ref{fig:3-address-code}. 

\subsubsection{Environment and Interpretation Functions} 
\label{sec:environment} 

The analysis interprets expressions and statements in an \emph{abstract} reference environment split into two components: $\Gamma$ and $\Gamma_0$. $\Gamma$ is a \emph{local} reference environment associated to the enclosing function. It is a map of (\ptt{id},\ptt{t}) tuples where \ptt{id} is a local and \ptt{t} is the \emph{analysis variable} representing the local. $\Gamma_0$ is the \emph{global} reference environment mapping identifiers for module-level constructs to analysis variables representing those constructs. In addition, there is an \emph{external} environment, $\Gamma_\mathit{ext}$, necessary for concrete evaluation. We analyze imports and group them into two categories: (1)~internal imports are relative imports (e.g., \mbox{\python{from ..metrics import get_scorer}}) and ones referencing the analysis package, and (2)~the remaining external imports (e.g., \python{import re}). 
External imports comprise $\Gamma_\mathit{ext}$.

The interpretation of a statement $\iComp{s}{\Gamma}$ returns a pair $(\Gamma', S)$ where $\Gamma'$ is the augmented local environment resulting from the interpretation of $s$ into 3-address code, and $S$ is \emph{the sequence of 3-address statements} corresponding to $s$. E.g., $\mathcal{I}(\ptt{x = y.f.g}, [(\ptt{y},\ptt{t1})])$ returns the augmented environment [(\ptt{x},\ptt{t2}), (\ptt{y},\ptt{t1})] and the following sequence of 3-address statements: \ptt{t4 = t1.f; t3 = t4.g; t2 = t3}. Here \ptt{t1} and \ptt{t2} are the analysis variables associated with local variables \ptt{y} and \ptt{x} respectively (in code examples we sometimes ignore analysis variables and use the source-level identifier directly). We leave $\Gamma_0$ and $\Gamma_\mathit{ext}$ implicit in the writeup as they are uniquely defined for each statement under interpretation.

The interpretation of an expression $\mathcal{I}(\pit{e}, \Gamma)$ returns a pair $(V,S)$ where $V$ is \emph{a set of analysis variables} and $S$ is the sequence of 3-address statements corresponding to \pit{e}. E.g., $\mathcal{I}(\ptt{y.f.g}, [(\ptt{y},\ptt{t1})])$ returns fresh variable \ptt{t3} and the sequence of 3-address statements \ptt{t4 = t1.f; t3 = t4.g}.


$\Phi$ is the environment of interpreted functions: a mapping from a function definition (Python AST FunctionDefs in the implementation) to a pair $(\Gamma,S)$, where $\Gamma$ is the local reference environment resulting from the translation of the function definition and $S$ is the sequence of 3-address statements corresponding to the function body.

%
%
The analysis uses the worklist algorithm in Figure~\ref{fig:worklist_algorithm}.

\begin{figure}
\begin{algorithmic}\small
\STATE \algComment{Initialize $\Gamma_0$:}
\STATE $\Gamma_0 = [], \Phi = \{\}$ 
\STATE {\bf for} $\langle\pymodule{\pit{M}}{\pit{i};\pit{s}}\rangle$ in package under analysis
\STATE $\quad$ {\bf for} $\langle\pyclass{C}{...}{...}\rangle$ in $\pit{s}$: $\Gamma_0 \leftarrow [(\pit{M.C},\ptt{t})]+\Gamma_0$, \ptt{t} is fresh
\STATE $\quad$ {\bf for} $\langle\pyfun{f}{...}{...}\rangle$ in $\pit{s}$: $\Gamma_0 \leftarrow [(\pit{M.f},\ptt{t})]+\Gamma_0$, \ptt{t} is fresh
\STATE $\quad$ \algComment{Imports \pyfrom{\pit{p}}{x'} \pkw{as} x are implicit assignments}
\STATE $\quad$ {\bf for} $\pyassign{\pit{x}}{...}$ in $\pit{s}$: $\Gamma_0 \leftarrow [(\pit{M.x},\ptt{t})]+\Gamma_0$, \ptt{t} is fresh
\STATE \algComment{Next, compute class hierarchy $H$ and MRO:}
\STATE $H \leftarrow \mathit{C3}(\Gamma_0)$ \algComment{$H$ maps $(\langle\pyclass{C}{...}{...}\rangle, \ptt{f})$ to $\langle\pyfun{f}{...}{...}\rangle$}
\STATE \algComment{Interpret \ptt{main} and add to worklist:}
\STATE $\Phi[\langle\pyfun{main}{...}{\pit{s}}\rangle] \leftarrow \mathcal{I}(\pit{s}, [])$
\STATE $W \leftarrow \{ \langle\pyfun{main}{...}{\pit{s}}\rangle \}$ \algComment{Entry point}
\STATE \algComment{Interpret module initializers and add to worklist:}
\STATE {\bf for} $\langle\pymodule{M}{\pit{...}}\rangle$ in package under analysis
\STATE  $\quad$ $\Phi[\langle\pyfun{M.module\_init}{...}{\pit{i};\pit{s}}\rangle] \leftarrow \mathcal{I}(\pit{i};\pit{s}, [])$ 
\STATE  $\quad$ $W \leftarrow \{ \langle\pyfun{M.module\_init}{...}{\pit{i};\pit{s}}\rangle \}$ \algComment{Entry points}
\STATE \algComment{Solve constraints in reachable functions:}
\STATE {\bf while} {$\textit{W} \neq \emptyset$}
\STATE $\quad \langle\pyfun{f}{...}{...}\rangle \leftarrow$ remove function from $\textit{W}$
\STATE $\quad$ {\bf for} $c$ in $\Phi[\langle\pyfun{f}{...}{...}\rangle][1]$: $\pit{W} \leftarrow \pit{W} + c.\mathit{solve}()$
\end{algorithmic}
\caption{\label{fig:worklist_algorithm}Worklist algorithm.}
\end{figure}


\subsubsection{Global Environment Initialization} 

The analysis first initializes a global environment $\Gamma_0$ that contains mappings for identifiers of module-level constructs, $M.f$ (functions), $M.C$ (classes), and $M.x$ (identifier definitions) to their corresponding analysis variables. The scope of these constructs spans the entire package, hence the analysis augments the environment and makes these constructs available during 3-address code generation. This is similar to the \emph{let rec} construct in functional programming which extends the scope for let-bound identifiers across all right-hand-side expressions.  Initialization does not interpret functions, classes and right-hand-side of assignments; these are interpreted during reachability analysis. 

\subsubsection{Class Hierarchy Analysis}

Next, the analysis computes the class hierarchy~$H$ using the C3 linearization algorithm for method resolution with multiple inheritance \cite{barrett_et_al_1996}. $H$ is a mapping from a pair of class definition $\langle\pyclass{C}{...}{...}\rangle$ and function name $f$ to the corresponding function definition $\langle\pyfun{f}{...}{...}\rangle$ resulting from a lookup in the \emph{Method Resolution Order} (MRO) of $C$. Additionally, the step creates a (\ptt{meta-cls}, $\langle\pyclass{C}{...}{...}\rangle$) object for each module-level class definition and associates the corresponding analysis variable $t$ to that object. In other words, this step creates an initial set of points-to edges which allows for data object creation during the second phase of the analysis.

\subsubsection{Iteration}
As it is customary for whole-program analysis, the analysis starts from a \ptt{main} function. It translates \ptt{main} and all module initializers into 3-address code (the calls to $\iComp{s}{[]}$) and adds them to the worklist. The analysis removes a function from the worklist and processes the 3-address statements $c$ for that function. The call $\mathit{c.solve()}$ solves the semantic constraints associated to $c$ which has side effects on the 3-address code environment $\Phi$ and points-to graph $\mathit{Pt}$. 
A call statement triggers interpretation and placement on the worklist of the callee function and it is an invariant that when a function is removed from the worklist, its three address code is available in $\Phi$. 

The call $\mathit{c.solve()}$ returns a minimal set of functions that are affected by solving constraint~$c$. For example, if $c$ is the 3-address code statement \ptt{t1 = t2} and solving it changes the points-to graph of \ptt{t1}, $\mathit{solve}$ returns the enclosing function of the statement. A change in the points-to graph due to \ptt{t1.f = t2} returns all functions in $\Phi$, as the effect of new objects added to $\mathit{Pt}(\ptt{o.f})$, where \ptt{o} is an object in the points-to set of \ptt{t1}, may propagate to arbitrary functions.

\begin{figure}
\[
\begin{array}{@{\hspace*{8mm}}lll}
    \iComp{\pypass}{\Gamma}
    &=&
    \iRet{\Gamma}{\set{}}
    \\[2em]
    \iComp{\pyassign{x}{e}}{\Gamma}
    &=&
    \begin{stack}
    (R,S) \leftarrow \iComp{e}{\Gamma}\\
    \text{if scope is}\ \pit{M.module\_init}:\\ 
    \quad \ptt{t} \leftarrow \pit{lookup}(M.x, \Gamma_0) \\
    \quad \iRet{\Gamma}{S \cup \set{\ptt{t} = t_e \mid t_e \in R}} \\
    \text{if}\ x \in \Gamma:\\
    \quad \ptt{t} \leftarrow \pit{lookup}(x, \Gamma) \\
    \quad \iRet{\Gamma}{S \cup \set{\ptt{t} = t_e \mid t_e \in R}} \\
    \text{else:}\\
    \quad \ptt{t} \leftarrow \ \text{fresh variable} \\
    \quad \Gamma' \leftarrow \env{(x,\ptt{t})} + \Gamma \\
    \quad \iRet{\Gamma'}{S \cup \set{\ptt{t} = t_e \mid t_e \in R}}
    \end{stack}
    \\[15em]
    \iComp{\pyseq{s_1}{s_2}}{\Gamma}
    &=&
    \begin{stack}
        (\Gamma_1, S_1) \leftarrow \iComp{s_1}{\Gamma}\\
        (\Gamma_2, S_2) \leftarrow \iComp{s_2}{\Gamma_1}\\
        \iRet
            {\Gamma_2}
            {S_1 \cup S_2}
    \end{stack}
    \\[5em]
    \iComp{\pyfor{x}{e}{s}}{\Gamma}
    &=& \iComp{\pyseq{\pyassign{x}{e}}{s}}{\Gamma}
    \\[2em]
    \iComp{\pyfun{f}{\pit{args}}{\pit{body}}}{\Gamma}
    &=&
    \begin{stack}
        \text{if scope is}\ \pit{M.module\_init}: \\  
        \quad \ptt{t} \leftarrow \pit{lookup}(M.f, \Gamma_0) \\
        \quad \iRet{\Gamma}{\set{\ptt{t} = (\ptt{meta-func},\ \langle\pyfun{f}{\pit{args}}{\pit{body}} \rangle)}} \\
        \text{else:}\\ 
        \quad \ptt{t} \leftarrow \text{fresh variable} \\
        \quad \Gamma' \leftarrow \env{(f, \ptt{t})} + \Gamma \\
        \quad \iRet
            {\Gamma'}
            {\set{\ptt{t} = (\ptt{meta-func},\ \langle\pyfun{f}{\pit{args}}{\pit{body}} \rangle)}}
    \end{stack}
    \\[9em]
    \iComp{\pyclass{C}{args}{body}}{\Gamma}
    &=&
    \iRet{\Gamma}{\set{}}
    \\[2em]
    \iComp{\pyother{s_1, ..., s_n}}{\Gamma}
    &=&
    \begin{stack}
    (\Gamma_1, S_1) \leftarrow \iComp{s_1}{\Gamma}\\
    ...\\
    (\Gamma_n, S_n) \leftarrow \iComp{s_n}{\Gamma_{n-1}}\\
    \iRet
        {\Gamma_n}
        {S_1 \cup ... \cup S_n}
    \end{stack}
\end{array}
\]
\caption{From Python statements to 3-address-code.
  Given an environment $\Gamma$, the interpretation function for a
  statement $\iComp{s}{\Gamma} = (\Gamma', S)$ returns an updated
  environment $\Gamma'$ and the 3-address code~$S$.
  $M$ is the enclosing module, and $\Gamma_0$ is the global environment.}
\label{fig:3-addr-stmt}
\end{figure}

\begin{figure}
\[
\begin{array}{@{}lll@{}}
\iComp{x}{\Gamma}
&=&
\begin{stack}
    \text{if}\ x \in \Gamma\!: \ \iRet{\set{\pit{lookup}(x, \Gamma)}}{\set{}}\\
    \text{if}\ M.x \in \Gamma_0\!: \ \iRet{\set{\pit{lookup}(M.x, \Gamma_0)}}{\set{}}\\
    \text{if}\ (\ptt{const},\mathit{ty},v) \leftarrow {\pit{eval}(x,\Gamma_\mathit{ext})}\!: \\ 
    \quad \iRet{\set{\pit{x}}}{\set{x = (\ptt{const},\mathit{ty},v)}}\\
    \text{else:}\ \iRet{\set{}}{\set{}}
\end{stack}
\\[6em]
\iComp{\pyattr{e}{f}}{\Gamma}
&=&
\begin{stack}
    \ptt{t} \leftarrow \ \text{fresh variable}\\
    \text{if}\ (\ptt{const},\mathit{ty},v) \leftarrow {\pit{eval}(\pyattr{e}{f},\Gamma_\mathit{ext})}\!: \\ 
    \quad \iRet{\set{\ptt{t}}}{\set{\ptt{t} = (\ptt{const},\mathit{ty},v)}}\\
    (V, S) \leftarrow \iComp{e}{\Gamma}\\
    \iRet
        {\set{\ptt{t}}}
        {S \cup \set{\ptt{t} = \pyattr{t_e}{f} \mid  t_e \in V}}
\end{stack}
\\[6em]
\iComp{\pysub{e_1}{e_2}}{\Gamma}
&=&
\begin{stack}
    \ptt{t} \leftarrow \ \text{fresh variable}\\
    \text{if}\ (\ptt{const},\mathit{ty},v) \leftarrow {\pit{eval}(\pysub{e_1}{e_2},\Gamma_\mathit{ext})}\!: \\ 
    \quad \iRet{\set{\ptt{t}}}{\set{\ptt{t} = (\ptt{const},\mathit{ty},v)}}\\
    (V_1, S_1) \leftarrow \iComp{e_1}{\Gamma}\\
    (V_2, S_2) \leftarrow \iComp{e_2}{\Gamma}\\
    \iRet
        {\set{\ptt{t}}}
        {S_1 \cup S_2 \cup \set{\ptt{t} = \pysub{t_1}{} \mid  t_1 \in V_1}}
\end{stack}
\\[7em]
\iComp{\pycall{e_1}{e_2}}{\Gamma}
&=&
\begin{stack}
    \ptt{t} \leftarrow \ \text{fresh variable}\\
    \text{if}\ (\ptt{const},\mathit{ty},v) \leftarrow {\pit{eval}(\pycall{e_1}{e_2},\Gamma_\mathit{ext})}\!: \\ 
    \quad \iRet{\set{\ptt{t}}}{\set{\ptt{t} = (\ptt{const},\mathit{ty},v)}}\\
    (V_1, S_1) \leftarrow \iComp{e_1}{\Gamma}\\
    (V_2, S_2) \leftarrow \iComp{e_2}{\Gamma}\\
    \iRet
        {\set{\ptt{t}}}
        {S_1 \cup S_2 \cup \set{\ptt{t} = \pycall{t_1}{t_2} \mid  t_1 \in V_1, t_2 \in V_2}}
\end{stack}
\\[7em]
\iComp{\pylist{e}}{\Gamma}
&=&
\begin{stack}
    \ptt{t} \leftarrow \ \text{fresh variable}\\
    \text{if}\ (\ptt{const},\mathit{ty},v) \leftarrow {\pit{eval}(\pylist{e},\Gamma_\mathit{ext})}\!: \\ 
    \quad \iRet{\set{\ptt{t}}}{\set{\ptt{t} = (\ptt{const},\mathit{ty},v)}}\\
    (V, S) \leftarrow \iComp{e}{\Gamma}\\
    S' \leftarrow \set{\ptt{t} = (\ptt{data},\ \langle \pkw{class}\ \ptt{list} \rangle)} \cup \set{\pysub{\ptt{t}}{} = t_e \mid  t_e \in V}\\
    \iRet
        {\set{\ptt{t}}}
        {S \cup S'}
\end{stack}
\\[7em]
\iComp{\pyother{e_1, ..., e_n}}{\Gamma}
&=&
\begin{stack}
    \text{if}\ (\ptt{const},\mathit{ty},v) \leftarrow {\pit{eval}(\pyother{e_1, ..., e_n},\Gamma_\mathit{ext})}\!: \\
    \quad \ptt{t} \leftarrow \ \text{fresh variable}\\
    \quad \iRet{\set{\ptt{t}}}{\set{\ptt{t} = (\ptt{const},\mathit{ty},v)}} \\
    \text{else:} \\
    \quad (V_1, S_1) \leftarrow \iComp{e_1}{\Gamma}\\
    \quad ... \\
    \quad (V_n, S_n) \leftarrow \iComp{e_n}{\Gamma}\\
    \quad  \iRet
        {V_1 \cup ... \cup V_n}
        {S_1 \cup ... \cup S_n}
\end{stack}
\\[9.4em]
\iComp{\pyas{\pyfrom{p}{x'}}{x}}{[]}
    &=&
    \begin{stack}
    M' \leftarrow \text{find module of}\ x'\\
    \ptt{t1} \leftarrow \pit{lookup}(M'.x', \Gamma_0)\\
    \ptt{t2} \leftarrow \pit{lookup}(M.x, \Gamma_0)\\
    \iRet{\set{}}{\set{\ptt{t2} = \ptt{t1}}}
    \end{stack}
\end{array}
\]
\caption{From Python expressions and imports to 3-address-code.
  Given an environment $\Gamma$, the interpretation function for an
  expression $\iComp{e}{\Gamma} = (V, S)$ returns a set of analysis
  variables $V$ and the 3-address code $S$.
  $M$ is the expression's enclosing module, and $\Gamma_\mathit{ext}$ is its 
  external environment. $\Gamma_0$ is the global environment.} 
\label{fig:3-addr-expr}
\end{figure}

\subsubsection{Interpretation of statements and expressions}
The definition of the interpretation function is presented in
Figures \ref{fig:3-addr-stmt} and~\ref{fig:3-addr-expr}. Interpretation of statements is in the abstract while interpretation of expressions weaves concrete and abstract evaluation. Below we highlight the most interesting points.

\paragraph{Statements}
Consider interpretation of assignment $\pyassign{x}{e}$. If the variable $x$ appears in a module initializer, we retrieve the analysis variable from $\Gamma_0$ (imports also create implicit global assignments). Otherwise, we only augment the local environment with a fresh variable if the variable $x$ is not in the current environment. 
We target flow-insensitive points-to analysis, and thus, a Python sequence \mbox{\python{x = 1;  x = "a"}} gives rise to 3-address code sequence \ptt{t1 = 1; t1 = "a"} and fails to distinguish that \python{x} is an integer at the first assignment and a string at the second.

The interpretation of a loop \pyfor{\pit{x}}{\pit{e}}{\pit{s}} reduces to a sequence of assignment \pyassign{\pit{x}}{\pit{e}} followed by \pit{s}. The assignment binds identifier \pit{x} before descending into the interpretation of \pit{s}.

A function definition in a module initializer has an entry \pit{M.f} in~$\Gamma_0$.
Otherwise, i.e., if this function is nested into another function, we augment the local environment with \pit{f} and return the augmented environment. Function definition gives rise to a ``new'' statement, assigning the abstract \ptt{meta-func} object to \ptt{t}. 
Class definitions have no effect during interpretation. 
Module-level class definitions are processed during class hierarchy analysis.

Note also that we ignore static reference environments for nested functions, which is unsound in general. A function value may flow to arbitrary points of the program and it is interpreted into 3-address code when it is called (detailed in calls in~\Cref{sec:constraint_resolution}); however, interpretation happens in the empty environment rather than the actual static reference environment and references coming from enclosing scopes evaluate to empty sets.


For uninterpreted statements, the algorithm descends into each sub-statement and extracts the corresponding 3-address code.
This does not ``glue'' components according to the statement's semantics. However, recursive descent processes all nested assignments and calls; it augments the environment and generates 3-address code that captures value flow. 

\paragraph{Expressions} 

Recall that a key feature of our analysis is concrete evaluation and that concrete evaluation happens both during 3-address code generation and constraint resolution. We employ the following heuristic for 3-address code generation, as the rules for expression interpretation in~\Cref{fig:3-addr-expr} detail. For simple expressions (i.e., variables), we first attempt resolution in the abstract environment: first $\Gamma$ then $\Gamma_0$. If it fails, we attempt concrete evaluation. For all other expressions (i.e., complex expressions), we first attempt concrete evaluation in $\Gamma_\mathit{ext}$, and if it fails to produce a concrete object we proceed with interpretation in the abstract. 


The interpretation of a variable (first rule in~\Cref{fig:3-addr-expr}) first searches the local environment~$\Gamma$ (i.e., the enclosing function), and then the global environment $\Gamma_0$ if the first lookup fails. If lookup in $\Gamma_0$ fails as well, it attempts concrete evaluation. If concrete evaluation succeeds and returns a concrete object, $\mathcal{I}(x,\Gamma)$ returns \pit{x} and an object creation statement. The last clause aims at built-in functions, e.g., \python{len} and enables reporting of built-in callees. The reader may have noticed here that we reuse the original variable name as an analysis variable, but create a new concrete instance of the built-in function object. Thus, there are as many concrete instances representing \python{len} as there are (static) calls to \python{len} in the code. Our implementation does not optimize this case yet, but it is trivial to get rid of such redundant concrete objects.


For remaining expressions, including \textit{Other}, the analysis first tries concrete evaluation, and if it succeeds, it returns the new analysis variable \pit{t} along with the assignment of the concrete object to \pit{t}. If an expression does not evaluate (which is the common case), interpretation continues recursively in the abstract environment. For example, \python{np.array(arg)} matches Call in~\Cref{fig:3-addr-expr}. Concrete evaluation fails with a NameError as \python{arg} is a local program variable; interpretation proceeds in the abstract. It recurses into Attribute \python{np.array} and this time it does evaluate to the numpy function. Interpretation of the call expression returns (\{\ptt{t1}\}, \{\ptt{t2 = (const, <... builtin ...>,...); t1 = t2(t3)}\}) where \ptt{t3} is the analysis variable for \python{arg}. 

Subscript expressions treat [] as a special field in the abstract, which is standard in points-to analysis. To interpret a list in the abstract, we create a new list object and generate subscript assignments to populate the list. Tuple, set, and dictionary are analogous. 

For uninterpreted expressions we return the union resulting from the interpretation of each sub-expression when concrete evaluation fails.

\paragraph{Import}
Finally, to interpret imports in module initializers (last rule of~\Cref{fig:3-addr-expr}), we first find the enclosing module $M'$ of imported construct $x'$, and then look up for the representative analysis variables \ptt{t1} and \ptt{t2} corresponding to the imported construct $M'.x'$ and the alias~$x$ in the current module $M$. Statement \ptt{t2 = t1} propagates the value (e.g., a function definition, a class definition) from module $M'$ to $M$. For example, consider statement \python{from cerberus.platform import \_get_args as get\_args} in module $M$. Lookup of \python{cerberus.platform.\_get_args} yields \ptt{t1} which points to the function def of \ptt{\_get\_args}. Lookup of $M$.\ptt{get\_args} yields \ptt{t2} and \ptt{t2 = t1} ensures the def propagates to \ptt{t2} and references to \ptt{get\_args} in $M$ find that function def. 

For simplicity, the rule treats $x'$ as a module-level construct, however, it can be a module. E.g., \python{errors} in \python{from cerberus import errors} is a module and code can access module-level constructs as in \python{errors.DocumentErrorTree}. We handle this case in the attribute rule: if lookup of $e$ in $\Gamma$ during $\mathcal{I}(e,\Gamma)$ yields no result, the analysis maps \pyattr{\pit{e}}{\pit{f}} to a module-level-construct identifier \pit{M.C}, \pit{M.f} or \pit{M.x}, and searches $\Gamma_0$ to retrieve the corresponding analysis variable (the step is not shown in the rule to avoid the complexity of lookup failure).

\subsection{Constraint Resolution}
\label{sec:constraint_resolution}

As mentioned earlier, the analysis maintains abstract objects and concrete objects. Abstract objects are explicitly grouped into \emph{data objects}, \emph{function objects}, and \emph{class object}:
\[
\begin{array}{ll}
\ptt{(data, $\langle\pyclass{C}{...}{...}\rangle$)} & \mathit{an~abstract~data~object} \\
\ptt{(meta-func, $\langle\pyfun{f}{...}{...}\rangle$)} & \mathit{an~abstract~function~object} \\
\ptt{(meta-cls, $\langle\pyclass{C}{...}{...}\rangle$)} & \mathit{an~abstract~class~object} \\
\ptt{(const, ty, v)} & \mathit{a~concrete~object} \\
\end{array}
\]

The 3-address statements are as specified in Figure~\ref{fig:3-address-code}. These statements induce constraints that populate a points-to graph $Pt$. The nodes in the points-to graph are analysis variables as well as objects $o$ of the above kinds. The edges represent the points-to relation. E.g., there could be an edge from variable $t$ to an object $o$, and this is denoted as $\{ o \} \subseteq \mathit{Pt}(t)$ or equivalently as $t \rightarrow o$. There could be a field edge indicating that field \pit{f} of $o_1$ points to $o_2$ and this is denoted as $\{ o_2 \} \subseteq \mathit{Pt}(o_1.\pit{f})$ or equivalently as $o_1 \stackrel{\pit{f}}{\rightarrow} o_2$. 

The rules for new assignment, copy propagation and field write are largely standard (except for their returns) and we elide them from the presentation. We elaborate on the rules for field read $\pyassign{t_1}{\pyattr{t_2}{f}}$ and function call $\pyassign{t_1}{\pycall{t_2}{t_3}}$ as they illustrate Python-specific semantics and concrete evaluation.

\subsubsection{Indirect read}

\begin{figure}\small
{$\mathit{solve}$ for $\pyassign{t_1}{\pyattr{t_2}{f}}$ in $\langle\pyfun{f'}{...}{...}\rangle$ with $\Gamma_\mathit{ext}:$}
\begin{algorithmic} 
\STATE {\bf for} $o \in \mathit{Pt}(t_2)$
\STATE $\quad$ {\bf case} $o$ {\bf of}
\STATE $\qquad \ptt{(data, $\langle\pyclass{C}{...}{...}\rangle$)} \rightarrow$ 
\STATE $\quad\quad\quad \langle\pyfun{f}{\ptt{self},p}{...}\rangle \leftarrow H[(\langle\pyclass{C}{...}{...}\rangle,\pit{f})]$
\STATE $\quad\quad\quad \mathit{Pt}(t_1) \leftarrow \mathit{Pt}(t_1) + \{\ptt{(meta-func, $\langle\pyfun{f}{o,p}{...}\rangle$)} \}$
\STATE $\qquad \ptt{(meta-cls, $\langle\pyclass{C}{...}{...}\rangle$)} \rightarrow$ 
\STATE $\quad\quad\quad \langle\pyfun{f}{\ptt{self},p}{...}\rangle \leftarrow H[(\langle\pyclass{C}{...}{...}\rangle,\pit{f})]$
\STATE $\quad\quad\quad \mathit{Pt}(t_1) \leftarrow \mathit{Pt}(t_1) + \{\ptt{(meta-func, $\langle\pyfun{f}{\ptt{self},p}{...}\rangle$)} \}$
\STATE $\qquad \ptt{(const, ...)} \rightarrow \mathit{Pt}(t_1) \leftarrow \mathit{Pt}(t_1) + \{\mathit{eval}(o.\pit{f},\Gamma_\mathit{ext})\}$
\STATE $\quad \mathit{Pt}(t_1) \leftarrow \mathit{Pt}(t_1) + \mathit{Pt}(o.\pit{f})$ \algComment{when $f$ is an object field, add its points-to set}
\STATE {\bf return} $\{\langle \pyfun{f'}{...}{...}\rangle\}$ if $\mathit{Pt}(t_1)$ changed else $\{\}$ 
\end{algorithmic}
\caption{Indirect read. Solving $\pyassign{t_1}{\pyattr{t_2}{f}}$ in enclosing function $\langle\pyfun{f'}{...}{...}\rangle$ with $\Gamma_\mathit{ext}$.}\label{fig:indirect_read}
\end{figure}

\noindent
The rule in Figure~\ref{fig:indirect_read} examines each object in the points-to set of receiver variable $t_2$ and does case-by-case analysis on the object type. E.g., if it is an abstract data object of some user-defined class, the analysis searches the class hierarchy (MRO) to determine the function referenced by $o.f$, forms the closure by binding \pit{self} to the receiver object $o$ and adds the closure object to the points-to set of left-hand-side $t_1$. If the object is a concrete object, the analysis evaluates the field access returning a new concrete object and adding it to the points-to set of $t_1$. Here, and in Calls, we restrict the number of times a statement is evaluated successfully. The restriction, currently set to 2, is not only an optimization but also necessary for termination, otherwise the analysis would keep creating distinct concrete objects at the same statement during fixpoint iteration.

The last line returns the enclosing function ${f'}$ to the worklist if there is a change to the points-to set of $t_1$, as the change may trigger changes to other points-to sets in ${f'}$.

\subsubsection{Call}

\begin{figure}[t]
{$\mathit{solve}$ for $\pyassign{t_1}{\pycall{t_2}{t_3}}$ in $\langle\pyfun{f'}{...}{...}\rangle$ with $\Gamma_\mathit{ext}:$}
{\small\begin{algorithmic} 
\STATE {\bf for} $o \in \mathit{Pt}(t_2)$
\STATE $\quad$ {\bf if} $o$ is an abstract object
\STATE $\qquad$ {\bf case} $o$ {\bf of}
\STATE $\quad\quad\quad \ptt{(data, $\langle \pyclass{C}{...}{...}\rangle$)} \rightarrow$ \algComment{call on data/instance} 
\STATE $\quad\quad\quad\quad \pit{callee} \leftarrow H[(\langle\pyclass{C}{...}{...}\rangle,\ptt{'\_\_call\_\_'})]$
\STATE $\quad\quad\quad\quad \pit{rcv} \leftarrow \{ o \}$
\STATE $\quad\qquad \ptt{(meta-cls, $\langle\pyclass{C}{...}{...}\rangle$)} \rightarrow$ \algComment{constructor call}
\STATE $\quad\quad\quad\quad \pit{callee} \leftarrow H[(\langle\pyclass{C}{...}{...}\rangle,\ptt{'\_\_init\_\_'})]$
\STATE $\quad\quad\quad\quad \pit{rcv} \leftarrow \{ (\ptt{data}, \langle\pyclass{C}{...}{...}\rangle)\}$ \algComment{new object}
\STATE $\quad\qquad (\ptt{meta-func}, \langle\pyfun{f}{o',\pit{p}}{\pit{s})}\rangle) \rightarrow$ \algComment{closure call}
\STATE $\quad\quad\quad\quad \pit{callee} \leftarrow \langle\pyfun{f}{\ptt{self},p}{\pit{s}\rangle)} $
\STATE $\quad\quad\quad\quad \pit{rcv} \leftarrow \{ o' \}$
\STATE $\quad\qquad (\ptt{meta-func}, \langle\pyfun{f}{\pit{p}}{\pit{s}}\rangle) \rightarrow$
\STATE $\quad\quad\quad\quad \pit{callee} \leftarrow \langle\pyfun{f}{\pit{p}}{\pit{s}}\rangle $ \algComment{the function def}
\STATE $\quad\quad\quad\quad \pit{rcv} \leftarrow \ptt{None}$
\STATE $\quad\quad$ $\langle\pyfun{f}{\pit{p}}{\pit{s}}\rangle = \pit{callee}$ \algComment{match and deconstruct callee}
\STATE $\qquad$ {\bf if} $\langle\pyfun{f}{\pit{p}}{\pit{s}}\rangle \notin \Phi$  \algComment{callee is not interpreted}
\STATE $\quad\quad\quad \ptt{t} \leftarrow$ fresh variable
\STATE $\quad\quad\quad \Phi[\langle \pyfun{f}{\pit{p}}{\pit{s}}\rangle] \leftarrow \mathcal{I}(\pit{s}, [(\pit{p},\ptt{t})])$ \algComment{env. includes \ptt{self} when \pit{f} is an instance function} 
\STATE $\quad\quad$ {\bf if} $\pit{rcv} \neq \ptt{None}$ \algComment{there is a receiver}
\STATE $\quad\quad\quad t_4 \leftarrow \Phi[\langle\pyfun{f}{\ptt{self},p}{s}\rangle][0][\ptt{self}]$ \algComment{retrieve analysis variable corresponding to \ptt{self}}
\STATE $\quad\quad\quad \mathit{Pt}(t_4) \leftarrow \mathit{Pt}(t_4) + \pit{rcv}$ \algComment{receiver to \ptt{self}}
\STATE $\quad\quad$ $t_5 \leftarrow \Phi[\langle\pyfun{f}{\pit{p}}{\pit{s}}\rangle][0][\pit{p}]$
\STATE $\quad\quad$ $\mathit{Pt}(t_5) \leftarrow \mathit{Pt}(t_5) + \mathit{Pt}(t_3)$ \algComment{actual to formal}
\STATE $\quad\quad$ $t_6 \leftarrow \Phi[\langle\pyfun{f}{\pit{p}}{s}\rangle][0][f\ptt{\_ret}]$ 
\STATE $\quad\quad$ $\mathit{Pt}(t_1) \leftarrow \mathit{Pt}(t_1) + \mathit{Pt}(t_6)$ \algComment{ret var to lhs of call}
\STATE $\quad$ {\bf else} \algComment{$o$ is a concrete object}
\STATE $\qquad$ {\bf for} $o_1$ in $\mathit{Pt}(t_3)$
\STATE $\quad\quad\quad$ {\bf if} $o_1$ is a concrete object 
\STATE $\quad\quad\quad\quad \mathit{Pt}(t_1) \leftarrow \mathit{Pt}(t_1) + \{\mathit{eval}(o(o_1),\Gamma_\mathit{ext})\}$
\STATE {\bf return} $\{\langle \pyfun{f}{...}{...}\rangle, \langle \pyfun{f'}{...}{...}\rangle\}$ if change else $\{\}$
\end{algorithmic}}
\caption{Call: solving for $\pyassign{t_1}{\pycall{t_2}{t_3}}$ in enclosing function $\langle\pyfun{f'}{...}{...}\rangle$ with $\Gamma_\mathit{ext}$.}
\label{fig:call}
\end{figure}

Figure~\ref{fig:call} details the rule for calls. There are two cases at the top level, an abstract object or a concrete object as function value. For an abstract object, we do case-by-case analysis. If $o$ is a data object, it queries the hierarchy to retrieve the corresponding \ptt{\_\_call\_\_} function --- this is the function that is being called. If it is a meta class object, this leads to the retrieval of the constructor. If it is a meta function, there are two cases: a closure where \ptt{self} is already bound to a receiver, or the function is just a value with a null reference environment. 

Once the analysis identifies the function to be called at this site, it checks if an interpretation of this function into 3-address code already exists. If it does not we interpret it. Notably, we interpret the AST of the function in the empty environment (i.e., only parameters are bound). This means that a first-class function does not carry its static reference environment and the analysis introduces unsoundness. While it is possible to extend the analysis with such bindings, this will complicate the code, while we believe in practice it will have limited impact (based on looking at Python code and our results). A notable departure from standard Java analysis is that there is no explicit \ptt{new A()} site. Different meta class objects may flow to receivers of calls accounting for data object creation (but the number of meta class objects is bounded and thus the number of data objects instantiated at the call is finite).

Once the callee function is interpreted, the analysis propagates points-to sets from actual arguments $t_3$ to formal parameters \pit{p} and return values to the left-hand-side of the call $t_1$. 
In case of a concrete receiver, if the analysis finds concrete arguments for all parameters, it executes the function.



\section{Results}
\label{sec:results}



We evaluate our Andersen-style points-to analysis, PoTo, on two clients:
type inference~(Section~\ref{sec:type_inference}) and
call graph construction~(Section~\ref{sec:call_graphs}).
However, points-to analysis has a wide variety of applications and we
envision other clients as well.
The analysis can be run on any Python package.
It starts at a provided entry function and computes a points-to graph 
containing information on reachable variables and their inferred types. 
To thoroughly analyze a library package, a set of entry functions are needed.
We follow DLInfer, a neural type inference for Python, and use the
same 10 Python packages from their experiment available with DLInfer's artifact.
The 10 packages range in size from 3,556 LOC to 285,515 LOC (see Table~\ref{tab:statistic}).
These packages contain rich test cases which suit our need of diverse entry points. 
We use each function in a package's test directory as an entry point.
For five of the packages (cerberus, mtgjson, pygal, sc2, and zfsp), we additionally 
create custom entry functions targeting remaining unreachable public functions. 
For the other five packages (anaconda, ansible, bokeh, invoke, and wemake\_python\_styleguide), we
only use default test suites.

\subsection{Type Inference}
\label{sec:type_inference}

To target unreachable methods for the purposes of type inference, we enhance the analysis.
We achieve this by performing a shallow analysis that collects built-in type 
information and type annotations from assignment and return statements.
For example, consider an assignment with a built-in type: 
\mbox{\python{rules = set(schema.get(field, ()))}.}
The shallow analysis infers that \python{rules} can be of type \python{set}.
This has impact when the method is unreachable from PoTo's entry points, 
as it creates a key and infers a type for the \python{rules} local variable. 
The final results, called PoTo+, are stored as a dictionary of \emph{keys} to their inferred types. 
A \emph{key} is a tuple of (module name, function name, variable name), describing the 
variable and its scoping information. Keys largely correspond to local variables (including arguments and returns) in a package and are 
an abstraction for flow-insensitive analysis, the target of our work.
The remainder of this section uses the terms keys and variables interchangeably. 


We compare the results of PoTo+ against four other type inference techniques.
Three of these are based on the recent neural type inference work DLInfer~\cite{yan_et_al_2023}:
DL-ST, DL-DY, DL-ML. We make use of the result files available with DLInfer's artifact~\cite{yanyan_yan_2023_7575545}. 
DL-ST is their ground truth information, which is a combination of running the
Pysonar2 static tool and extracting type information~\cite{pysonar2}. 
DL-DY is a set of dynamic type information obtained from executing the test suites. 
This dynamic set contains only variables whose type can be collected only this way, 
meaning that this is the set difference of the actual dynamic run and DL-ST set~\cite{yan_et_al_2023}.
Lastly, DL-ML is the result of DLInfer's machine-learning approach.
We aggregate each DLInfer result in the same way as our analysis, which 
is a dictionary of keys to their types.
We choose DLInfer for several reasons: (1) DLInfer is recent work in a top conference, (2)~it compares with several state-of-the-art deep-learning techniques: Type4Py~\cite{mir_et_al_2022}, Typilus~\cite{allamanis_et_al_2020}, PYInfer~\cite{Cui_et_al_2021}, and DeepTyper~\cite{Hellendoorn_et_al_2018}, (3) DLInfer infers types for local variables, not only parameters and returns, and (4) its artifact~\cite{yanyan_yan_2023_7575545} includes full Python packages along with analysis results allowing for a comparison over the same code base.

In addition to DLInfer, we also compare our result to Pytype, a prominent static 
type checking and type inference tool~\cite{pytype_2016}.
To get type information for all variables, we instrument the package by inserting 
Pytype's command \python{reveal_type(var)} for each local variable at the end of a function, and run Pytype on each file in the package directory. 
The types are then collected and combined to be the inferred result of Pytype.
This process takes time to complete but only needs to be done once.
This is a departure from standard use of Pytype as baseline for type inference work, which uses Pytype's result on parameters and 
returns only (e.g.~\cite{peng_et_al_2022}); 
Pytype can infer types for local variables as well and we make use of this in our comparison.

\medskip

Our evaluation considers 4 research questions:
\begin{description}
  \item[RQ1:] How high is the coverage of PoTo+ compared to other type inference techniques?
  \item[RQ2:] To what extent are the types from PoTo+ equivalent to those from other techniques?
  \item[RQ3:] If the types do not match those from other techniques, which one is correct?
  \item[RQ4:] Does the time to run PoTo+ scale well and how does it compare to Pytype?
\end{description}

To answer RQ1, we measure the percentage of total keys in a package for which our analysis reports types against the
percentage of total keys for which the four other type inference techniques report types (more detail on methodology in Section~\ref{sec:rq1}). 
Our analysis collects types for a \emph{larger percentage of total keys} compared to the four other techniques.
To answer RQ2, we compare our inferred types to those inferred by each
of the four other techniques, measuring the equivalence in term of
total match, partial match, and mismatch (more detail in Section~\ref{sec:rq2}).
Our inferred types \emph{largely match with Pytype's}. 
While there is some match with DLInfer, there is disagreement in many keys.

To answer RQ3, we inspect a sample of mismatches according to the result from RQ2 for each pair comparisons:
PoTo+ vs. Pytype, PoTo+ vs. DL-*.
We found that in the few cases where there is a mismatch with Pytype, Pytype is correct in all cases. 
A mismatch with DL-* is nearly always a correct result by PoTo but an incorrect result by DL-*. 
The conclusion from RQ1-RQ3 is that \emph{traditional techniques outperform a state-of-the-art neural technique} 
for the task of type inference.
%
Lastly, for RQ4, we compare the total time to run PoTo (with dozens of entry functions for the smaller packages and thousands for the larger ones) to the time to run Pytype to collect reveal-type information. On all but one package, ansible, PoTo outperforms Pytype significantly. 

\subsubsection{Coverage (RQ1)}
\label{sec:rq1}

\begin{table}
\centering
\caption{Statistics for the dataset.}
{\small\begin{tabular}{@{}lrrrrr@{}}
\toprule
\multirow{2}{*}{\textsc{Package}} & \multicolumn{2}{c}{\textsc{Files}} & \multicolumn{2}{c}{\textsc{LOC}}       & \multirow{2}{*}{\textsc{Keys}} \\ \cmidrule(lr){2-3} \cmidrule(lr){4-5}
                                   & Total     & Test Dir    & Total       & Test Dir     &                             \\ \midrule
cerberus                           & 45                 & 32                   & 6,694                & 3,011                 & 966                         \\
mtgjson                            & 54                 & 2                    & 6,912                & 58                    & 1,378                       \\
pygal                              & 78                 & 24                   & 13,780               & 3,208                 & 2,439                       \\
sc2                                & 69                 & 6                    & 11,205               & 603                   & 5,970                       \\
zsfp                               & 54                 & 7                    & 3,556                & 207                   & 1,345                       \\
anaconda                           & 370                & 9                    & 90,207               & 587                   & 21,183                      \\
ansible                            & 1,445              & 961                  & 285,515              & 174,296               & 22,346                      \\
bokeh                              & 1,133              & 280                  & 131,931              & 43,316                & 14,978                      \\
invoke                             & 133                & 65                   & 26,159               & 9,878                 & 3,474                       \\
wemake                             & 403                & 291                  & 55,841               & 32,798                & 3,184                       \\ \bottomrule
\end{tabular}}


\label{tab:statistic}
\end{table}

Table~\ref{tab:statistic} shows the statistic of all 10 packages. 
We are interested in inferring types for the core package and exclude test files from our reports. 
The total keys include all variables, function arguments, and function return types. 
They serve as an upper limit of possible variables (or keys) for each package.

\begin{figure*}[!t]
\includegraphics[width=\textwidth,trim={10pt 60pt 60pt 120pt},clip]{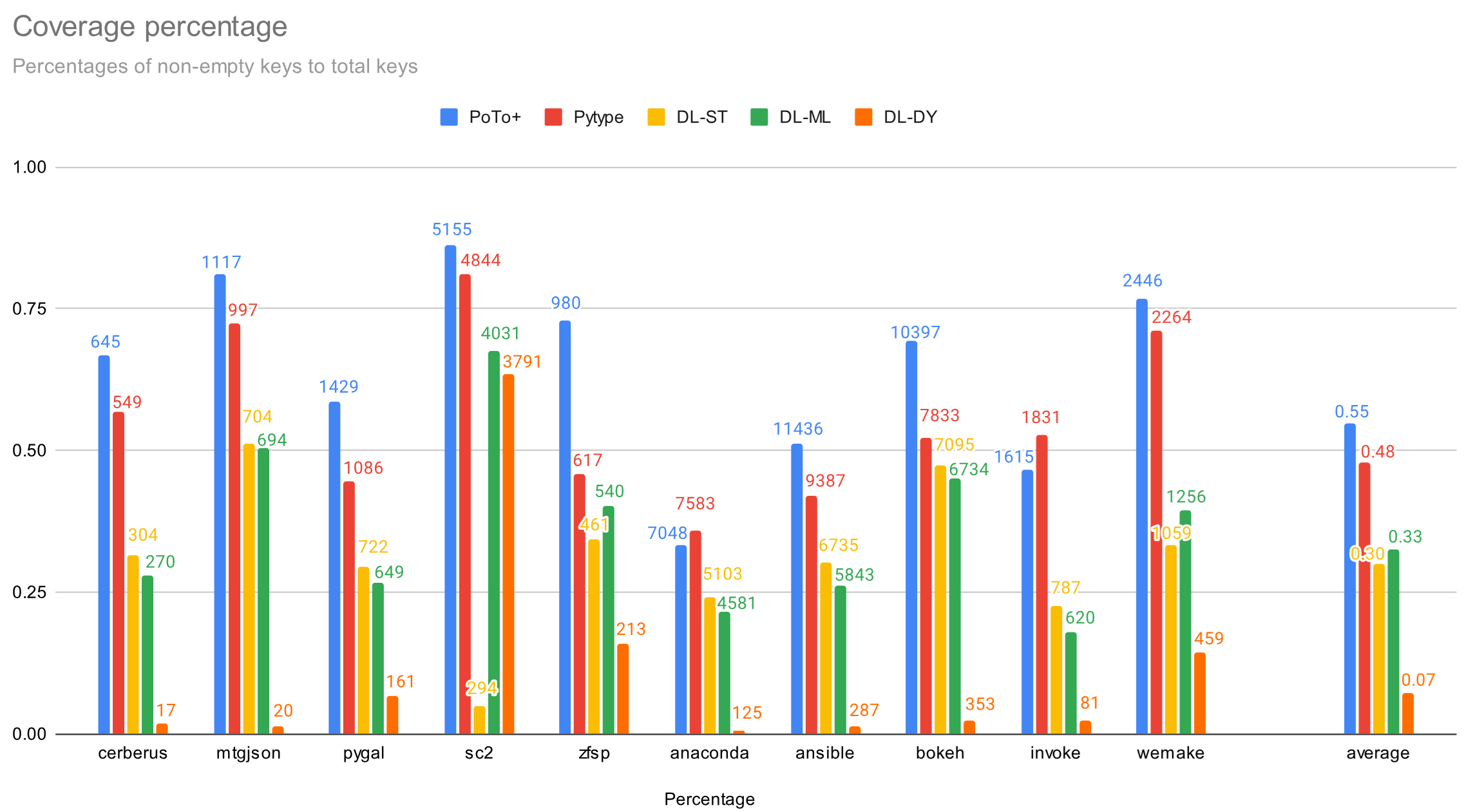}
\caption{Coverage percentages of non-empty keys to total keys (RQ1).}
\label{fig:coverage}
\end{figure*}

Figure~\ref{fig:coverage} shows the percentages of non-empty keys to the total keys 
for each package. 
Non-empty keys are ones for which the analysis (ours or other techniques) gives type information; 
empty keys are variables in total keys for which the analysis gives no information. 
In the case of Pytype, a key is designated as empty if its inferred
type is the trivial \python{Any} type.
The numbers above each bar are the numbers of non-empty keys.
For the five packages (cerberus, mtgjson, pygal, sc2, and zfsp) where we add custom entry functions, 
PoTo+ covers more non-empty keys than other techniques, followed closely by Pytype.
For the remaining packages where we only use default test suites,
we have slightly worse coverage than Pytype on anaconda and invoke, reflecting worse test coverage
by the underlying test suites. PoTo+ still has the highest average coverage percentage. 

\subsubsection{Equivalence (RQ2)}
\label{sec:rq2}

\newcommand*{\eqTrimmed}[2]{\includegraphics[width=#1,trim={5pt 32pt 5pt 45pt},clip]{#2}}

\begin{figure}[!t]
  \centerline{\includegraphics[width=0.9\textwidth,trim={15pt 20pt 15pt 20pt},clip]{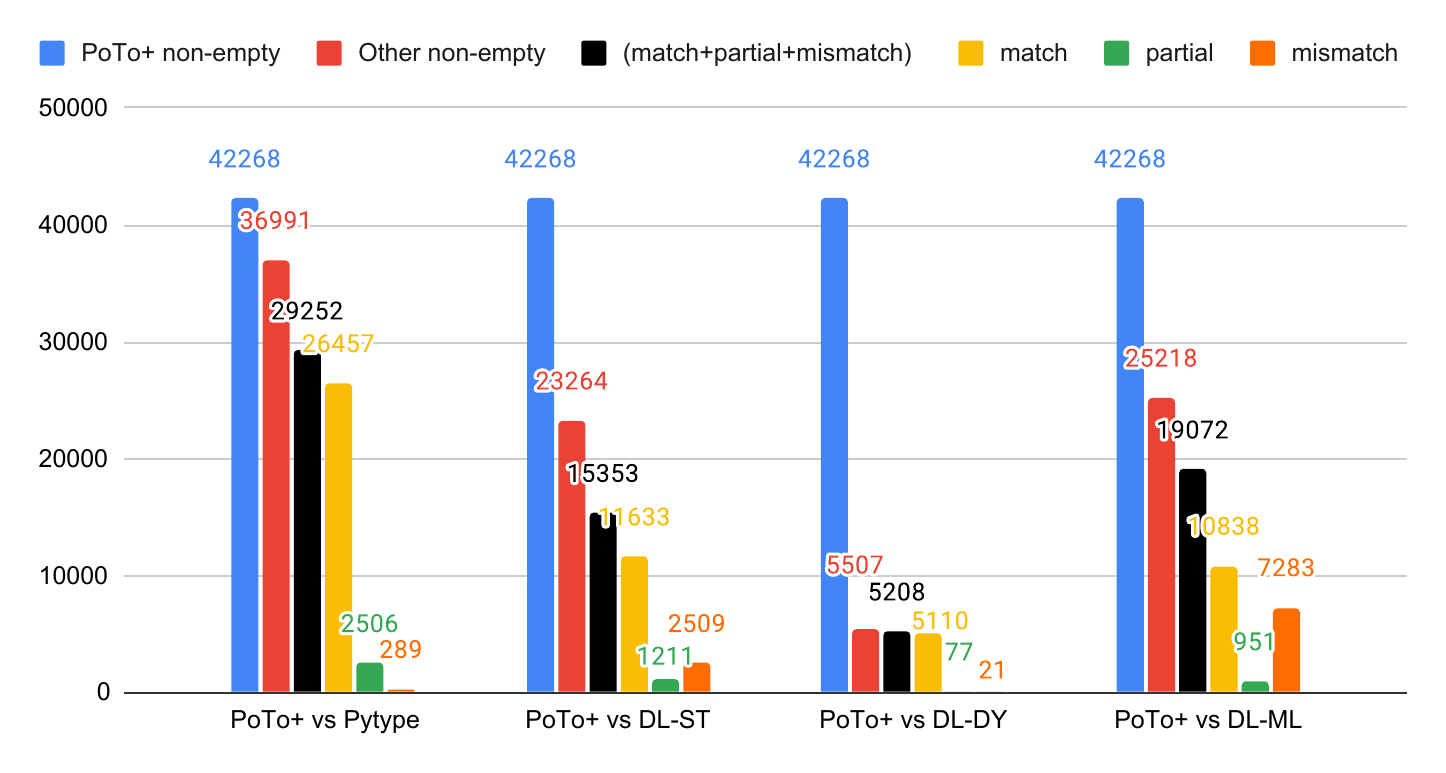}}
  \caption{\label{fig:equivalence_average}Equivalence comparison between PoTo+ and other type inference techniques (RQ2). Black bars (sums) show the number of keys for which both techniques report types. Numbers are summation of all 10 packages.} 
\end{figure}

Next, we measure the similarity of inferred types.
For each variable in a package, our analysis and the other type inference
techniques can either have information on this variable, and the information is the set of its 
inferred types, or have no information, i.e., an empty set.

For built-in containers such as \python{dict}, \python{list}, \python{set}, and \python{tuple}, we compare only the top-level 
part and deem that they match if they are the same container types.
The main reason for this shallow comparison is to facilitate processing and comparison 
as each type inference technique infers and reports types in different forms. 
For example, DLInfer reports only the type of a container, e.g., \python{dict}.
Pytype reports parametric information, e.g., \python{Dict[str, int]}, but not always, 
while PoTo+'s inferred types come from abstract and concrete objects, e.g., 
\python{\{'rename\_handler': <class 'int'>\}}. 
PoTo can collect parametric information, but it requires significant processing and raises issues on 
reporting types for polymorphic containers. 

Matching is automatic but some comparisons require manual verification because of rendering of PoTo concrete objects. 
For example, PoTo+'s \python{1970-01-01} matches with DL-ST's \python{datetime.date} type. 
Another example is a match between PoTo+'s \mbox{\python{<function <lambda>}} \python{at 0x109297520>} 
and Pytype's \python{Callable[[Any],Any]}.


We are interested in 3 groups of equivalence: total match, partial match, and mismatch, where partial match 
is when the two sets of types have non-empty intersection and mismatch is an empty intersection. 
Figure~\ref{fig:equivalence_average} shows these as summation across the 10 Python packages.
In addition, Figure~\ref{fig:equivalence_average} shows bars with the total number of non-empty 
keys by PoTo+ next to bars with the total number of non-empty keys by the other technique; for convenience 
it draws a bar that sums match, partial match and mismatch, i.e., the set of keys for which both techniques 
report a type.
On all Python packages, PoTo+ shares the most non-empty keys with Pytype, and has high
numbers of total matching. DL-ST and DL-ML have some matching and partial matching with our analysis, but also 
contain many keys that are mismatches. The discrepancy is discussed in Section~\ref{sec:rq3}.
For the sc2 package, almost all common keys between PoTo+ and DL-ML are mismatches.
The mismatches are due to simple assignments of class attributes in /sc2/ids/ directory, 
e.g. \python{NULL = 0}, \python{RADAR25 = 1}, \python{TAUNTB = 2}.
Our analysis, Pytype, and DL-DY correctly infer the types as integer, but DL-ML labels all 
of them as \python{num} which is incorrect as it is not a type. 
There are 3,791 such assignments out of 3,984 common keys between PoTo+ and DL-ML in the sc2 package.
Lastly, DL-DY has low coverage of types information, and most of them are a total 
match with our analysis. 

Figure~\ref{fig:equivalence_average} shows that PoTo+ reports a larger number of types compared to the other techniques; for example, 
for Pytype, there are 29,252 keys for which both report type information (the black bar) and there are 13,016 keys for which PoTo+ reports a type but Pytype does not. In contrast there are 7,739 keys for which Pytype reports a type but PoTo+ is empty. Appendix~\ref{sec:equivalence_average_all} includes per benchmark versions of this figure.

\subsubsection{Correctness (RQ3)}
\label{sec:rq3}

\begin{figure}
\centerline{\includegraphics[width=0.75\textwidth,trim=10pt 20pt 10pt 50pt,clip=true]{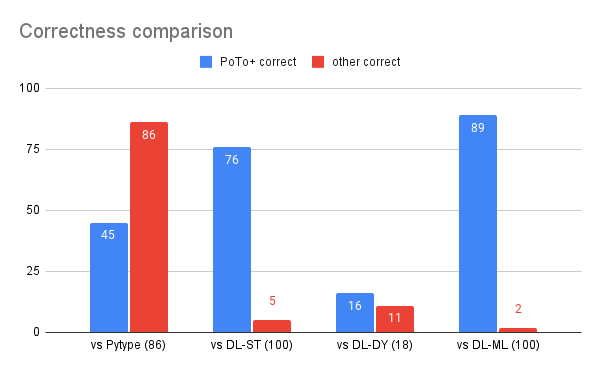}}
\caption{Correctness comparison between PoTo+ and other techniques (RQ3). Numbers in parentheses are \# of samples.
This figure zooms in on a sample of the mismatched keys from Figure~\ref{fig:equivalence_average}; for Pytype there are 
289 mismatched keys which is less than 1\% of all variables.}
\label{fig:correctness}
\end{figure}


To carry out the comparison we manually examine a sample of 10 non-match keys for each package and each pair of comparisons, i.e., PoTo+ vs. Pytype, PoTo+ vs. DL-ST, PoTo+ vs. DL-DY, and PoTo+ vs. DL-ML. In cases where there are fewer than 10 non-match keys, we exhaustively examine all of them, (e.g., there are only 8 non-matches between PoTo+ and Pytype for pygal, so there are 8 instead of 10 keys for pygal). 
There are only 18 total non-matches keys for PoTo+ vs. DL-DY. For the remaining 3 techniques, the total is slightly under 300 samples. Results are shown in Figure~\ref{fig:correctness}.

\paragraph{PoTo+ vs. Pytype} Comparison with Pytype shows that a mismatch usually means that PoTo+ is incorrect while Pytype is correct. 
Of the 86 pairs we examined, PoTo+ is correct 45 times, while Pytype is correct 86 times, because Pytype handles a larger set of Python features precisely, while PoTo defaults to Other, thus polluting points-to sets. For example, PoTo does handle list and dictionary comprehensions precisely, but defaults to Other on set comprehensions. Consider the example:
\python{rampPoints = \{ p for p in rampDict if rampDict[p]\}}.
Recall that handling of Other propagates points-to sets of \python{p}, \python{rampDict} and \python{rampDict[p]} into \python{rampPoints}. This leads to PoTo+ reporting type \python{[dict, bool]} as  \python{p}'s set is empty and \python{rampDict}'s set contains a single dictionary object of \python{Point} object keys, and bool values (it is straightforward to add handling of set comprehensions and other features to PoTo to improve matching).

For the majority of cases where both are correct, PoTo+ reports a concrete type while Pytype reports a type parameter. For example, PoTo+ reports correctly \python{[dict, None]}, i.e., an optional dictionary for the return type of \python{\_normalize\_purge\_unknown} in the cerberus package; while Pytype reports type variable \python{[\_T0]} and this is correct with respect to Pytype's type system which allows for polymorphic functions.  

\paragraph{PoTo+ vs. DL-ML, DL-ST and DL-DY} Of the 100 pairs we examine for PoTo+ vs. DL-ML, PoTo+ is correct 89 times, while DL-ML is correct 2 times. We mark PoTo+ incorrect conservatively --- essentially, when the result is partially correct but not sound and not complete, we mark it incorrect. One common source of incorrect result for DL-ML was the \python{num} types mentioned earlier. Another source is usage of the function name instead of the function's return type at call assignments. Consider the following example from package pygal: \mbox{\python{background = lighten('#e6e7e9',7)}.}
DL-ML reports type \python{[lighten]} for \python{background}, while PoTo+ correctly reports \python{[str, tuple]}. 
Yet another source of incorrect results is over-reliance on \python{dict} types.
For example, in the wemake package:
\mbox{\python{AnyFunctionDef=Union[ast.FunctionDef,ast.AsyncFunctionDef]}},
PoTo+ correctly infers type \python{[typing.Union[...]]} for \python{AnyFunctionDef} (via concrete evaluation). DL-ML infers \python{[dict]}.

Of the 100 samples we examine for PoTo+ vs. DL-ST, PoTo+ is correct 84 time and DL-ST is correct 5 times. There are no \python{num} types, as they are correctly assigned \python{int}; however, we observe similar patterns of function name instead of return type, and over-reliance on~\python{dict}.

There are only 18 total mismatches for PoTo+ vs. DL-DY; 8 from ansible and 10 from bokeh. 
PoTo+ is correct 16 times and DL-DY is correct 11 times. 
Most cases are from accessing a dictionary or calling a function such as \python{children = kwargs.get('children')} or \python{typ = type(obj)}. 
On 5 occasions, DL-DY infers types that appear wrong such as \python{type}. 
















    


    
    

    
    
      
\subsubsection{Scalability of PoTo (RQ4)}
\label{sec:rq4}      

\begin{table}
\centering
\caption{Running times for PoTo and Pytype (RQ4).}
{\small
\begin{tabular}{@{}lrr@{}}
\toprule
\textsc{Package} & \textsc{PoTo} & \textsc{Pytype} \\ \midrule
cerberus         & 51s           & 4m22s           \\
mtgjson          & 1m39s         & 15m39s          \\
pygal            & 1m48s         & 1h00m17s        \\
sc2              & 52s           & 40m29s          \\
zsfp             & 11s           & 7m28s           \\ \bottomrule
\end{tabular}
\hspace*{10mm}
\begin{tabular}{@{}lrr@{}}
\toprule
\textsc{Package} & \textsc{PoTo} & \textsc{Pytype} \\ \midrule
anaconda         & 1m59s         & 1h21m40s        \\
ansible          & 6h23m07s      & 6h14m59s        \\
bokeh            & 1h41m16s      & 10h15m41s       \\
invoke           & 1m32s         & 20m45s          \\
wemake           & 3m25s         & 9m17s           \\ \bottomrule
\end{tabular}}
\label{tab:running_time}
\end{table}

Table~\ref{tab:running_time} show the running times of PoTo (total time for thousands of entry functions) and of Pytype. We run on a commodity Mac book Pro with 2.4 GHz 8-Core Intel Core i9 and 32 GB of Memory (one of our development machines). The ``+'' phase (type inference) is instantaneous and type aggregation and processing are negligible. We do not include these in the timing. In both cases, execution is IO-dominated, as PoTo writes points-to results into pkl files and Pytype's reveal-type reveals types by printing special error message.

PoTo runs in 1--3 minutes for all but the two largest packages, bokeh and ansible. Pytype is significantly more expensive ranging between 4 and 81 minutes on those packages. Without the reveal-type instrumentation Pytype runs 30\% faster than with; thus, its underlying static analysis is expensive and PoTo still outperforms Pytype. 
PoTo runs slower than Pytype on ansible because ansible has nearly 500 test file (we filtered out ones that do not reference ansible packages from the 961 original ones to speed up testing) and many hit the same bottleneck of ansible code in the points-to analysis. 

\subsection{Call Graphs}
\label{sec:call_graphs}

To strengthen the case for applicability of PoTo, we evaluate it on another classical client of points-to analysis: call graph construction. We call the client PoToCG.
We compare PoToCG to PyCG, the leading Python call graph construction analysis~\cite{salis_et_al_2021}. Unfortunately, PoToCG and PyCG are two different analyses, with different goals, and comparing them is non-trivial. PoToCG is a standard whole-program reachability analysis~\cite{bacon_sweeny_1996,rountev_et_al_2001,li_et_al_2018},
whose goal is as follows: given a \emph{main function} and a package, compute the smallest, i.e., most precise, call graph of reachable functions.  PyCG, on the other hand, takes a set of files as input and computes a call graph that aims to include every call site in these files. Both graphs are unsound, by virtue of being analyses of Python, and both techniques aim to minimize the number of missing call graph nodes and edges. 
Nevertheless, PyCG is still the best Python call-graph analysis to compare to
representing the state-of-the-art, and we made the comparison as
apples-to-apples as possible modulo the different nature of the two
approaches.
We do not expand PoToCG to unreachable functions for this comparison, because such an expansion would entail a brand new call graph analysis that is antithetical to the reachability analysis inherent in Andersen-based call graph construction. Instead, we report on the number of reachable functions in PoToCG compared to PyCG, then we zoom in on the PoToCG-reachable functions and compare the quality of the PoToCG call graph for those functions compared to PyCG.

\medskip

The research questions we address are as follows:
\begin{description} 
\item[RQ5:] How does PoToCG compare to PyCG in terms of reachable functions?
\item[RQ6:] For the PoToCG-reachable functions, how \emph{complete} and how \emph{precise} is the PoToCG call graph compared to PyCG?

\end{description}

We construct the PoToCG call graph by traversing call 3-address statements \ptt{x = y(z)} in PoTo reachable functions and examining the points-to set of \ptt{y}. Meta-func objects represent abstract functions and lead to callees in the package. Concrete objects lead to callees that are either built-in functions (e.g., \python{len}) or external library functions (e.g., \python{re.sub}).

Consider function \python{UnconcernedValidator.\_\_init\_\_} in benchmark cerberus containing the following statement: 
\python{self.document\_error\_tree = errors.DocumentErrorTree()}.
It gives rise to a corresponding 3-address-code call statement \ptt{x = y(z)}. Analysis variable \ptt{y}'s points-to set contains meta-func abstract object \python{cerberus.errors.DocumentErrorTree.\_\_init\_\_}. Thus, we record a call graph edge from caller \python{UnconcernedValidator.\_\_init\_\_}  to callee function \python{cerberus.errors.DocumentErrorTree.\_\_init\_\_} (this edge is missing from the PyCG results). PoToCG handles calls to built-in functions as well by recording the concrete object corresponding to the built-in function, e.g., \python{len}, \python{int}, etc. PyCG captures built-in callees as well. We aggregate results across all test cases, as in the type inference client. 

Unfortunately, PyCG ran only on the five smaller packages ---  cerberus, pygal, mtgjson, sc2, and invoke. 
It did not terminate on any of the remaining packages. (PoToCG terminates on them.)
To run PyCG, all files in a package are passed to PyCG as entry points. 
The five packages that terminated took 30 seconds to a few minutes, 
which are about the same as PoToCG as shown in Table~\ref{tab:running_time}.
The other packages timed out after two hours per package.


\begin{table}
\centering
\caption{Function coverage of PoToCG and PyCG (RQ5). * indicates that the number is unusual due to non-existent callees, i.e., callees that do not correspond to function definitions.}
\label{tab:functions}
\vspace*{-2mm}
\begin{tikzpicture}
\node (table) {\small
\begin{tabular}{@{}lrrrrr@{}}
\toprule
\textsc{Functions}                                & cerberus & pygal & mtgjson & sc2 & invoke \\ \midrule
Total functions                                  & 200               & 354            & 237              & 567          & 782             \\
~~~Funcs without a call stmt         & 38                & 83             & 32               & 256          & 111             \\
~~~Funcs with call stmt              & 162               & 271            & 205              & 311          & 671             \\ \midrule
Funcs in PoToCG                                    & 91                & 218            & 211              & 266          & 99              \\
Funcs in PyCG                                    & 161               & 816\textsuperscript{*} & 259              & 412          & 738             \\
Funcs in PoToCG but not in PyCG                    & 5                 & 26             & 0                & 15           & 2               \\
Funcs in PyCG but not in PoToCG                    & 75                & 624            & 48               & 161          & 641             \\ \midrule
Funcs in both. Same \# of edges                  & 37                & 88             & 86               & 168          & 50              \\
~~~Exactly the same set of edges     & 29                & 71             & 57               & 123          & 37              \\
~~~Different set of edges            & 8                 & 17             & 29               & 45           & 13              \\ 
Funcs in both. \#Edges: PoToCG \textgreater{} PyCG & 43                & 54             & 76               & 67           & 33              \\
Funcs in both. \#Edges: PyCG \textgreater{} PoToCG & 6                 & 50             & 49               & 16           & 14              \\ \bottomrule
\end{tabular}};
\draw [white!50!black,ultra thick]
  ($(table.south west) !0.! (table.north west)$)
  rectangle 
  ($(table.south east) !.23! (table.north east)$);
\end{tikzpicture}
\end{table}

Table~\ref{tab:functions} shows function coverage of PoToCG and PyCG (RQ5). 
The top three rows provide the raw statistic of each package we produced by scanning its Python AST. 
For the result of both techniques (row 4 and 5), we are interested in functions that have at least one outgoing edge or one incoming edge in their respective call graphs, meaning that we exclude isolated empty functions. 
Except for invoke, which relies entirely on its test suite (we did not write additional tests for it), PoToCG achieves relatively good coverage: 
49\% for cerberus, 72\% for pygal, 91\% for mtgjson, and 77\% for sc2. 
We measure coverages by percentages of functions in PoToCG that have call statements, to total functions with call statements.
PyCG numbers for pygal and mtgjson 
exceed the upper bound of total functions, because PyCG reports caller and callee nodes that cannot be mapped to actual function definitions, e.g., \python{pygal.etree.etree.fromstring.findall.node.serie.get.split.append}.
PyCG includes each module as a node as it can potentially have call edges. 
PyCG also maps functions in \ptt{\_\_init\_\_.py} files differently. 
For example, function \python{safe\_load()} in \ptt{invoke/vendor/yaml3/\_\_init\_\_.py} has its path as 
\python{invoke.vendor.yaml3.safe\_load}. 
We adjust PoToCG to accommodate these cases.


\begin{table}
\centering
\caption{Edges in PoToCG reachable functions (RQ6).}
\label{tab:edges}
\vspace*{-2mm}
\begin{tabular}{@{}lrrrrr@{}}
\toprule
\textsc{Edges}                                & cerberus & pygal & mtgjson & sc2 & invoke \\ \midrule
True positive PoToCG & \textbf{339} & \textbf{599} & \textbf{679} & \textbf{438} & \textbf{260} \\
True positive PyCG & 192 & 398 & 550 & 351 & 220 \\ \midrule
False positive PoToCG & \textbf{1} & \textbf{0} & \textbf{0} & \textbf{0} & \textbf{0}\\
False positive PyCG & 12 & 813 & 116 & 4 & 6\\ \bottomrule
\end{tabular}
\end{table}

Table~\ref{tab:edges} zooms in on the functions that are both in PoToCG and PyCG for which the two techniques report \emph{different sets of callee functions} (RQ6). These are the bottom three lines in Table~\ref{tab:functions} surrounded with a solid rectangle. We examined all the edges in these functions. We classified a caller-callee edge as \emph{true positive} when reaching a specific call statement in the caller would indisputably lead to the callee for some receiver (when there is a receiver); this is the case for the \python{DocumentErrorTree} constructor example above. We classified an edge as \emph{false positive} otherwise. In practice, essentially all false positives were in PyCG where the reported callee could not be mapped to a function definition in the package or in an external library. One can see in Table~\ref{tab:edges} that PoToCG reports considerably better call graphs for the functions it reaches --- it reports a much larger number of true positive callees compared to PyCG as well as virtually no false positives, while PyCG reports a large number of false positives for two of the benchmarks. Therefore, PoTo's call graph is more complete and more precise for the functions it reaches. We observed that when reported callees could be mapped to function definitions, both PoToCG and PyCG's callees were true positives. The reason why PyCG had a large number of false positives was that it reported many callees that did not correspond to function definitions.

We consider cases that highlight PoToCG's better recall (i.e., soundness) and better precision. PoToCG's better recall is due to its flow analysis as in the following example:

\centerline{\python{validator = self.\_get\_child\_validator(..); mapping[field] = validator.normalized(..)}}

\noindent
In this example, the analysis must infer that a \python{Validator} object flows to the \python{validator} variable. PoTo infers that the analysis variable corresponding to \python{validator} points to a Validator object and correctly infers a call edge to \python{cerberus.base.UnconcernedValidator.normalized}. In contrast, PyCG leaves this call empty. 

Another reason for PoToCG's better coverage is concrete evaluation. Consider

\centerline{\python{result = validator.normalized(...); mapping[field] = value\_type(result.values())}}

\noindent
In the above code, an empty concrete dictionary flows to the analysis variable corresponding to \python{result} and concrete evaluation deduces that the \python{values} member function is called. In contrast, PyCG does not infer a callee at this call. 

As mentioned previously, PyCG reports hundreds of non-existent callees as in the following example: \python{pygal.util.decorate.serie.node}. Here, \python{decorate} is a function in the \ptt{util} module. We surmise PyCG infers \python{decorate} to return an object that has a field \python{serie}; however, it is not clear how that access path corresponds to an actual function definition in the pygal package. Another example is \python{pygal.graph.public.PublicApi.svg.serie.node}. We observed large numbers of such false positive callees in the pygal and mtgjson benchmarks but nearly none in the remaining benchmarks. In contrast, PoToCG reports were precise --- if the call is reached then the reported function will be called. 

Overall, we observed that recall (i.e., soundness) rather than precision is a more pressing issue for all analyses. PoToCG has better recall overall for the functions it reaches as detailed in Table~\ref{tab:edges}, however, in the absence of ground truth it is difficult to judge how complete PoTo's call graph is. We did not observe precision issues in either analysis results, other than PyCG reporting non-existent callees. We further examined several applications and their hierarchies and we could not easily identify a call site that would have benefitted from a context- or object-sensitive analysis. For example, in pygal, a charting library and one of the larger applications in the suite, classes encapsulated predominantly string and numerical values, as opposed to complex objects; superclasses contained nearly all the functionality; and there were no setter and getter functions. As a specific case, one class hierarchy had root \python{class Style} which defined several class-level string and numerical fields representing colors and fonts. It also defined a function \python{get\_colors} which formatted color strings to interact with CSS. Subclasses of \python{Style} added a small number of string or numerical class-level fields and defined no functions.
Appendix~\ref{sec:difficult_example} gives an example where PoTo shows a clear advantage over an object-sensitive analysis. We conjecture that Python applications use class hierarchies differently from Java although further studies are needed. We consider PoTo a necessary baseline. Improved precision, possibly via context-sensitive, is a promising direction for future work in Python program analysis.


\section{Threats to Validity}\label{sec:threats}

Section~\ref{sec:rq3} discusses correctness comparison that depends on limited samples out of thousands of mismatches. 
To mitigate this, we collect all 18 mismatches against DL-DY, plus 10 samples per packages for all packages across 3 other techniques, 
resulting in total of 300 hand-labeled examples with broad coverage.


The analysis relies on unit tests for entry points, thus presuming the existence of test suites and good test coverage of the test suites. To increase coverage, we added custom entry-points for 5 of the packages; notably, coverage remains robust even for those large packages where we used only the existing test suites.

Lastly, PoTo is a hybrid analysis that alternates between concrete and abstract evaluation, and it employs semantic choices to deal with Python's complexity. This makes the analysis unsound by design, but it is the typical kind of trade-off made by most real-world static analysis~\cite{livshits_et_al_2015}.
Our results with two client analyses show that PoTo compares favorably against respective state-of-the-art techniques, thus validating its trade-offs.

\section{Related Work}\label{sec:related}

This section discusses prior work related to each of the three
contributions stated in Section~\ref{sec:introduction}.

\emph{Points-to analysis.}
At its core, PoTo is an implementation of Andersen's points-to
analysis~\cite{andersen_1994}, but unlike the original, it works for
Python and is hybridized with concrete evaluation.
The only other points-to analysis for Python we have found is in
Scalpel~\cite{li_wang_quan_2022}; however, Scalpel's analysis is not
based on Andersen's, does not use concrete evaluation, and the paper
lacks empirical results.
Few static analyses have been shown to work on
real-world Python programs, including PyCG~\cite{salis_et_al_2021}
(which finds call graphs) and Tree-sitter~\cite{clem_thomson_2022}
(which is limited to syntactic queries)~\cite{clem_thomson_2022}.
Neither PyCG nor Tree-sitter does points-to analysis, nor do they use
concrete evaluation.
This paper empirically compares PoToCG against PyCG.

Outside of Python, work on points-to analysis and related alias and value-flow analysis dates decades back. Chatterjee et al.~\cite{Chatterjee_et_al_1999} describe a context-sensitive points-to analysis for C++, Rountev et al.~\cite{rountev_et_al_2001} present an Andersen-style analysis for Java, and F{\"a}hndrich et al.~\cite{Fahndrich_et_al_2000} define a context-sensitive points-to analysis for C, among many other works. Siu and Xue~\cite{Siu_and_Xue_2016} and Shi et al.~\cite{Shi_et_al_2018} present scalable value-flow analysis over LLVM IR. Jang and Kwang~\cite{Jang_and_Kwang_2009} define the first Andersen-like points-to analysis for JavaScript and Sridharan et al.~\cite{Sridharan_et_al_2012} improve Andersen's analysis with better reasoning about property accesses. These analyses work over established 3-address code IR in LLVM, Wala, and Soot. In contrast, we needed to construct 3-address code for points-to analysis from Python AST constructs.



Recent works on points-to analysis for Java (e.g., \cite{Lu_and_Xue_2019}, \cite{Jeon_and_Oh_2022}) focus on context- and object-sensitivity to improve precision. Our analysis is a context-insensitive and flow-insensitive baseline; we envision that the 3-address code and hybridization will serve as a foundation for building context- and flow-sensitive analysis for Python in the future.

\emph{Hybridization weaving concrete and abstract evaluation.}
PoTo uses concrete evaluation to solve the problem of analyzing Python
programs that use external libraries.
Two other works hybridize abstract (i.e.\ static) with concrete
(i.e.\ dynamic) analysis for Python:
PyCT~\cite{Chen_et_al_2021} (which does concolic testing) and
Rak-amnouy\-kit et al.'s analysis~\cite{rakamnouykit_et_al_2024}
(which finds weakest preconditions).
Neither is based on points-to analysis, nor has been used for type inference.

Looking back outside of Python, the idea of combining static and dynamic analyses has a long history in program analysis of Java and JavaScript. Hirzel et al.~\cite{hirzel_diwan_hind_2004} hybridize Andersen's analysis with dynamic analysis to handle Java's dynamic class loading. Grech et al.~\cite{Grech_et_al_2017} define a hybrid analysis for Java that takes whole heap snapshots and integrates them into the analysis. Wei and Ryder~\cite{Wei_and_Ryder_2013} present blended analysis for JavaScript that collects traces of execution then uses static analysis to reason about flow of tainted values. Tripp et al.'s analysis for JavaScript~\cite{Tripp_et_al_2014} addresses the problem of modeling the DOM; it runs a dynamic analysis to concretize the external environment, followed by a static analysis in the concrete environment. Similarly, Laursen et al.~\cite{laursen_et_al_2024} do a dynamic pre-analysis followed by a static analysis that uses runtime results; they target property access in JavaScript, a notorious issue in JavaScript codes~\cite{chakraborty_et_al_2022}. Our analysis differs from all these works as it targets Python. It presents a novel blend of static and dynamic interpretation --- it weaves concrete evaluation in both 3-address code generation and constraint resolution and elevates concrete objects to first-class status in the analysis. Park et al.~\cite{Park_et_al_2021}'s analysis of JavaScript does concrete execution of selected functions using carefully constructed objects; this is similar to our usage, however, their goal is to improve precision, while our goal is to improve soundness.

We view Toman and Grossman's Concerto~\cite{Toman_and_Grossman_2019} as most closely related to our work. Concerto is a combined concrete and abstract interpretation framework. It considers concrete state~(for framework code) and abstract state (for application code) and allows calls from one domain into the other. Concerto is a theoretical framework, while we focus on Python and develop 3-address code generation and a specific analysis, Andersen's points-to analysis.




\emph{Type inference and call graph analysis for Python.}
PYInfer~\cite{Cui_et_al_2021} and DeepTyper~\cite{Hellendoorn_et_al_2018}
use deep learning for Python type inference, and several other works
combine deep learning with simple static
analysis~\cite{allamanis_et_al_2020,mir_et_al_2022,peng_et_al_2022,pradel_et_al_2020,xu_et_al_2016,yan_et_al_2023}.
This paper empirically compares PoTo+ against the latest of those,
DLInfer~\cite{yan_et_al_2023}, chosen because it compares with several
earlier works and its artifact~\cite{yanyan_yan_2023_7575545}
has points-to sets for 10~real-world Python programs.
A handful of other works use static analysis for Python type
inference~\cite{fritz_hage_2017,pytype_2016,hassan_et_al_2018,maia_moreira_reis_2011}.
Unlike PoTo+, none of these use points-to analysis nor concrete
evaluation.
In addition to DLInfer, this paper also empirically compares PoTo+
against Pytype~\cite{pytype_2016}, which is based on static analysis.
We chose Pytype as a second baseline because it is widely used in
practice and, like PoTo+, handles real-world Python
programs~\cite{rakamnouykit_et_al_2020}.
Even though call-graph analysis is a popular client for points-to
analysis in other languages, for Python, it has received little
attention from the academic community.
The main work is PyCG~\cite{salis_et_al_2021}, and thus, we chose it
as a baseline.

Instead of hybridizing static analysis with concrete evaluation,
several works hybridize static analysis with machine-learning~(ML).
Xu et al.\ present a static type inference for Python augmented with
ML for guessing types based on variable names~\cite{xu_et_al_2016}.
Typilus uses static analysis to build a Python program dependency
graph, then uses a graph neural network for type
inference~\cite{allamanis_et_al_2020}.
TypeWriter performs deep-learning (DL) type inference, then uses a
static type checker to repair hallucinations~\cite{pradel_et_al_2020}.
Type4Py~\cite{mir_et_al_2022},
HiTyper~\cite{peng_et_al_2022}, and DLInfer~\cite{yan_et_al_2023} are
primarly DL type inferences for Python, assisted by simple static
analyses.

\section{Conclusions}\label{sec:conclusions}

This paper presents PoTo, the first Andersen-style points-to analysis
for Python.
PoTo works on real-world Python programs, which use dynamic features
as well as external packages for which source code is often missing
(and which often involve non-Python code).
To handle external packages, PoTo introduces a novel hybridization of
Andersen's static analysis with dynamic concrete evaluation.
In terms of clients, this paper presents PoTo+, a type inference built upon
PoTo.
While several recent papers explore deep learning for Python type
inference, our results indicate that (at least as of now) static
analysis solves this problem with superior coverage and correctness.
This paper also presents
PoToCG, a call-graph analysis built upon PoTo
that performs better than the state-of-the-art for Python.

\bibliography{bibfile}

\newpage
\appendix
\section{Appendix}
\label{sec:supplement}

\subsection{Per-application Equivalence Comparisons including Bars for Non-Empty Variables}\label{sec:equivalence_average_all}

\begin{figure}[h!]
    \includegraphics[width=0.49\textwidth,trim={15pt 5pt 15pt 20pt},clip]{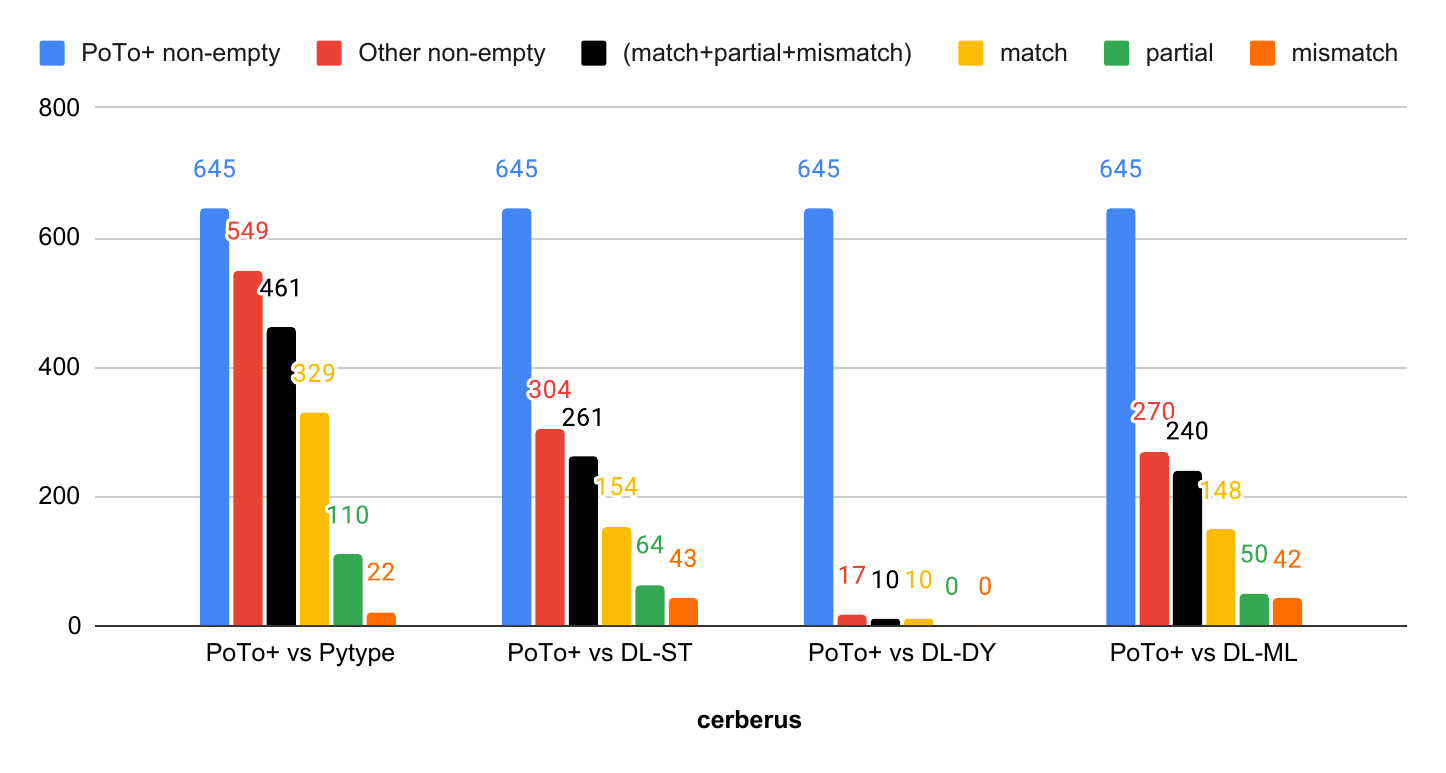}
    \includegraphics[width=0.49\textwidth,trim={15pt 5pt 15pt 20pt},clip]{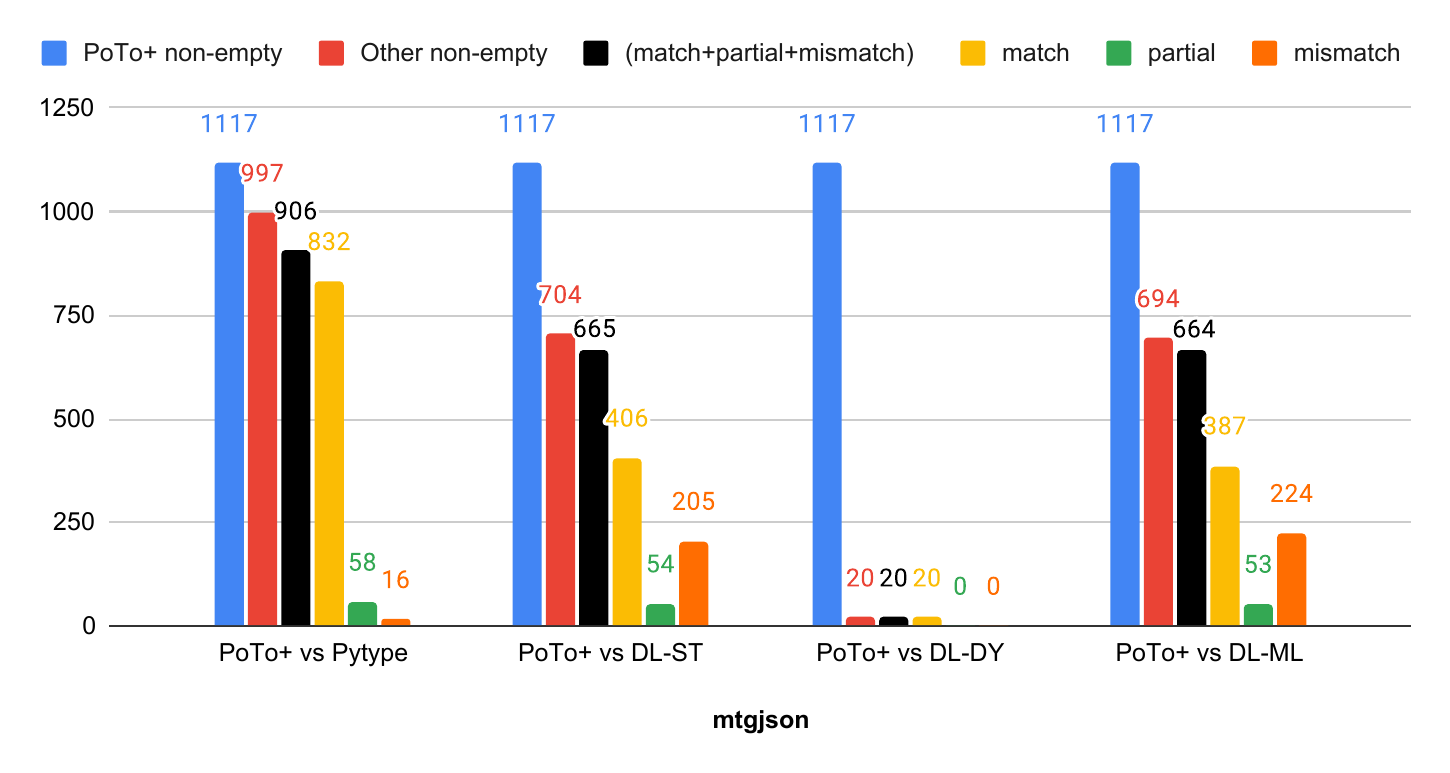}
    \includegraphics[width=0.49\textwidth,trim={15pt 5pt 15pt 20pt},clip]{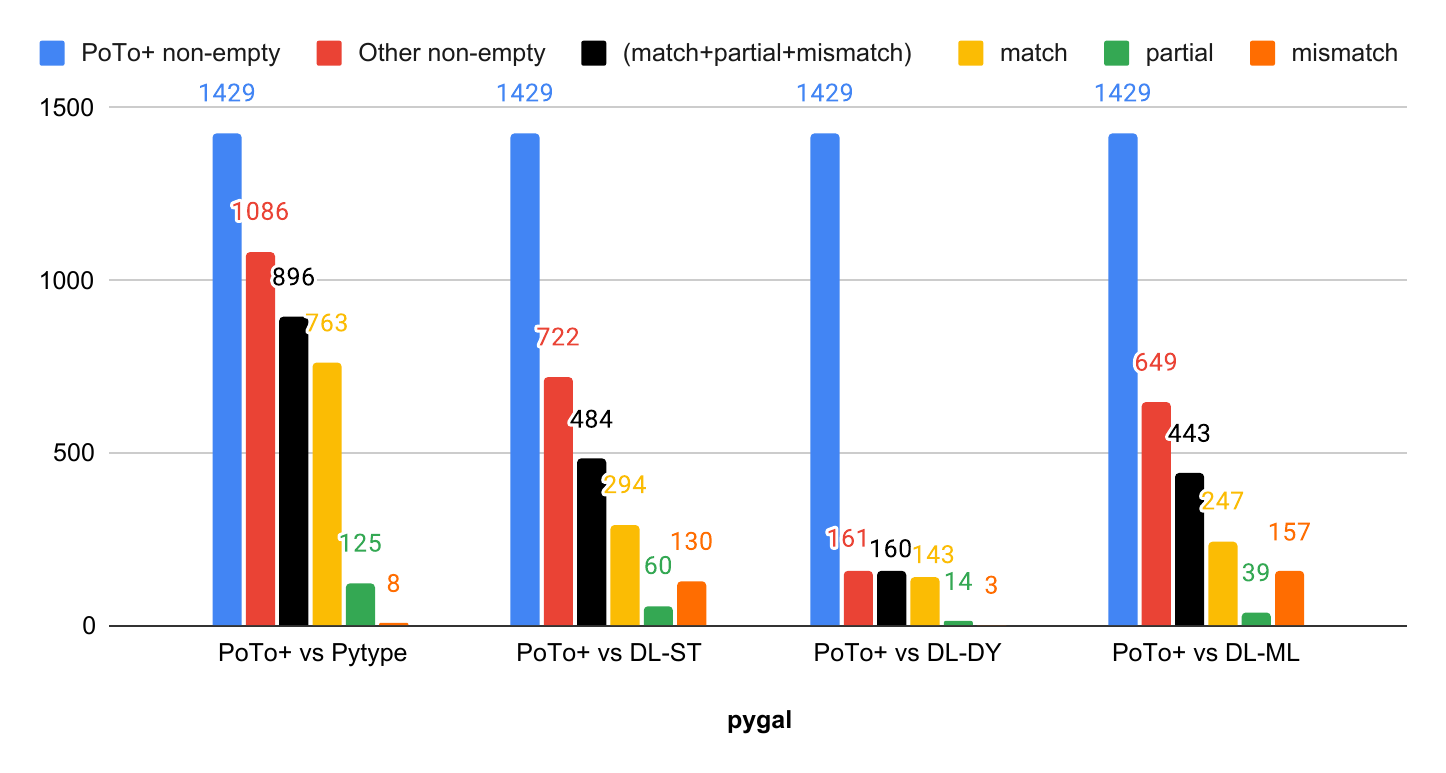}
    \includegraphics[width=0.49\textwidth,trim={15pt 5pt 15pt 20pt},clip]{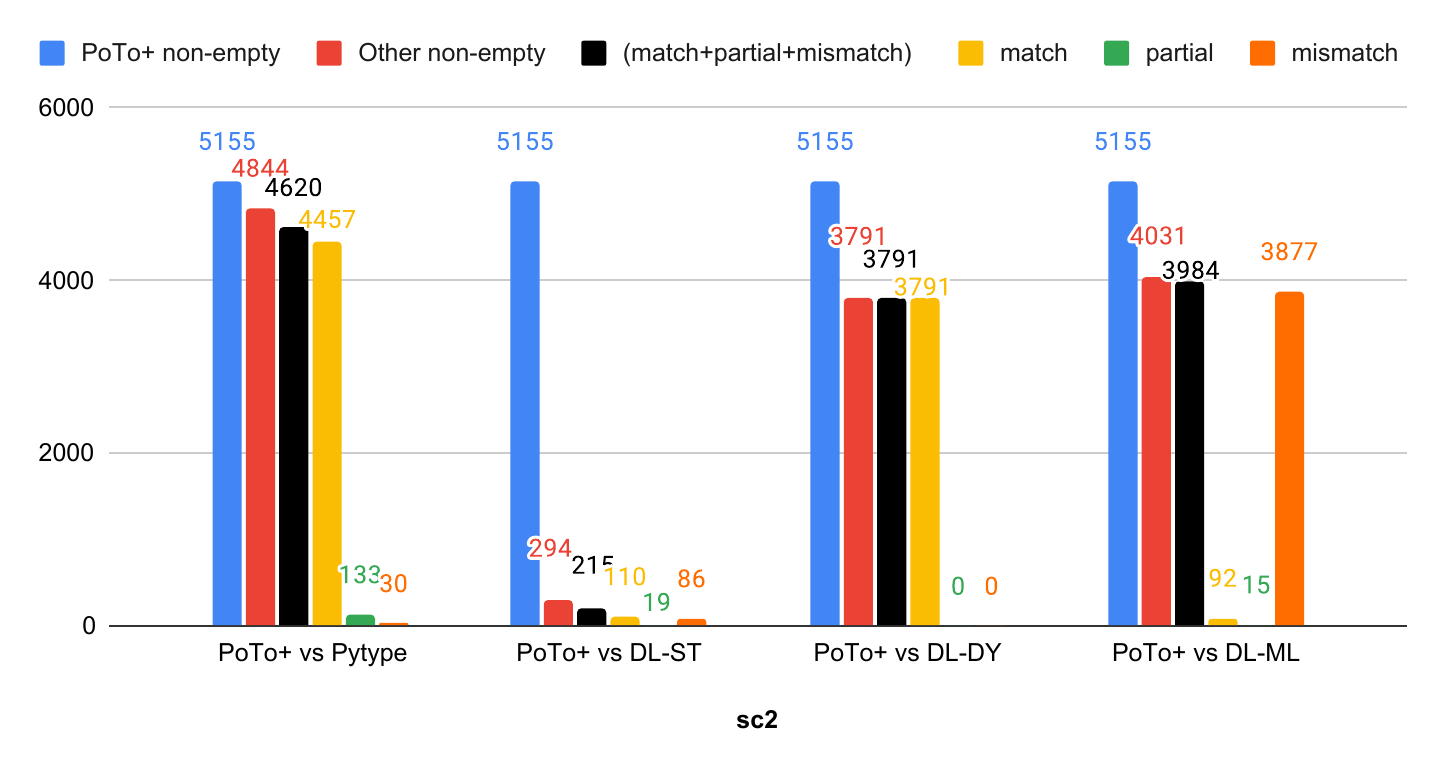}
    \includegraphics[width=0.49\textwidth,trim={15pt 5pt 15pt 20pt},clip]{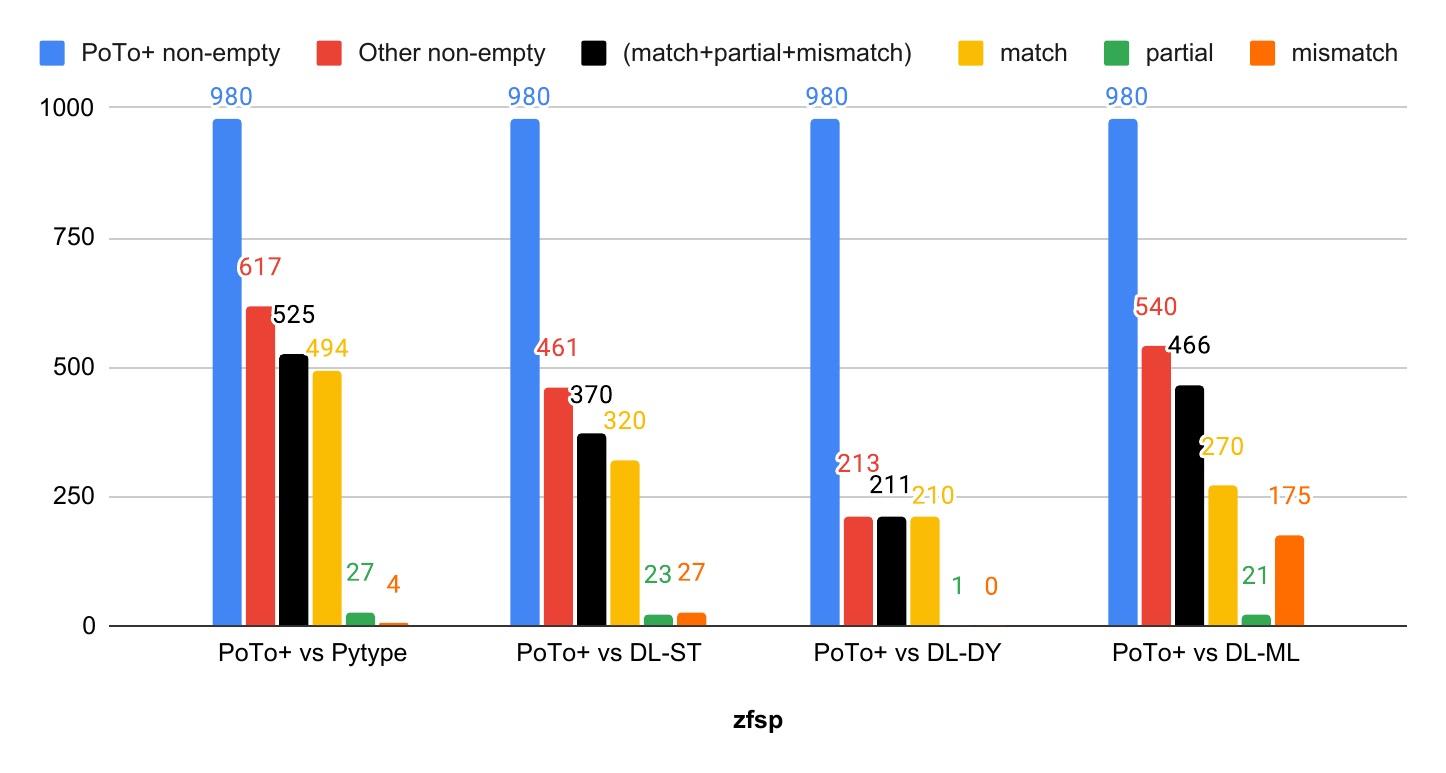}
    \includegraphics[width=0.49\textwidth,trim={15pt 5pt 15pt 20pt},clip]{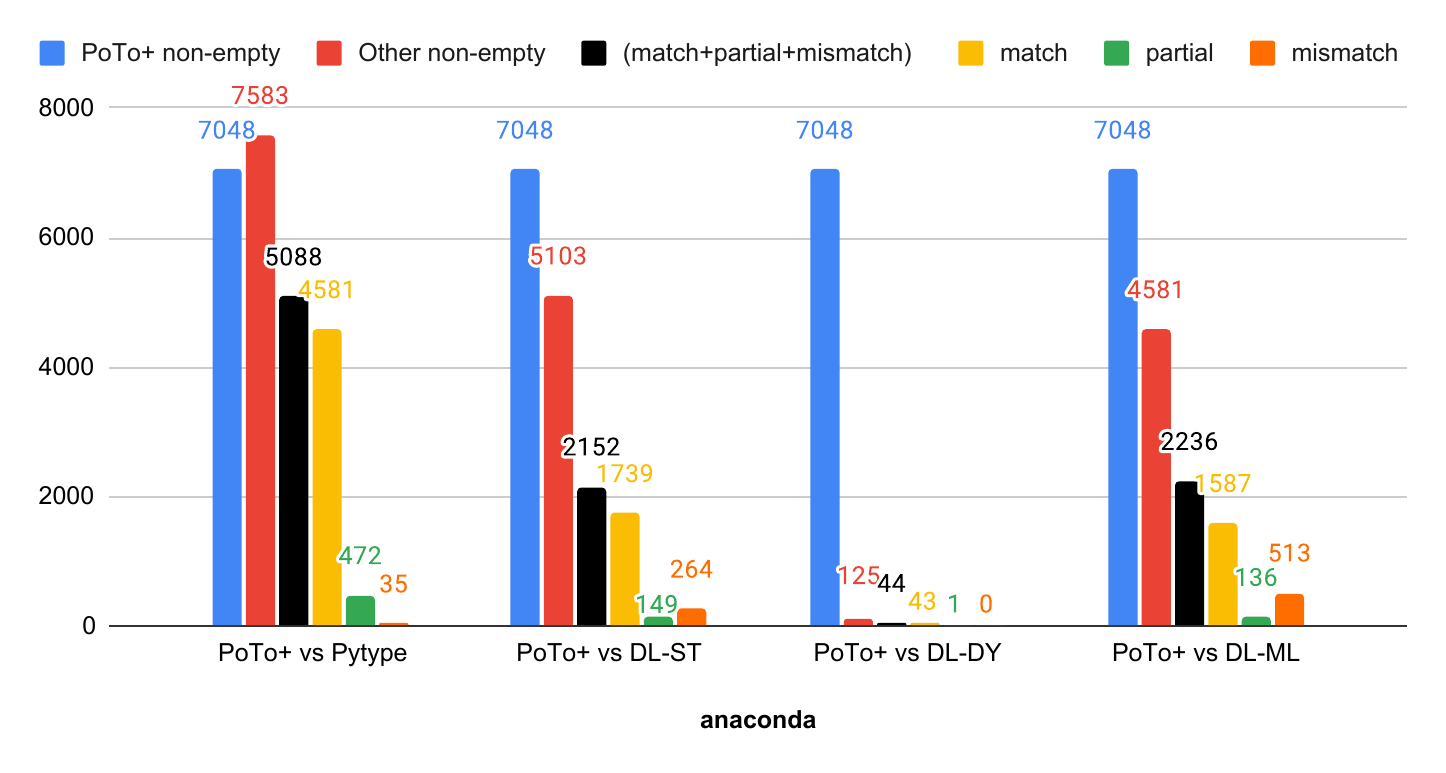}
    \includegraphics[width=0.49\textwidth,trim={15pt 5pt 15pt 20pt},clip]{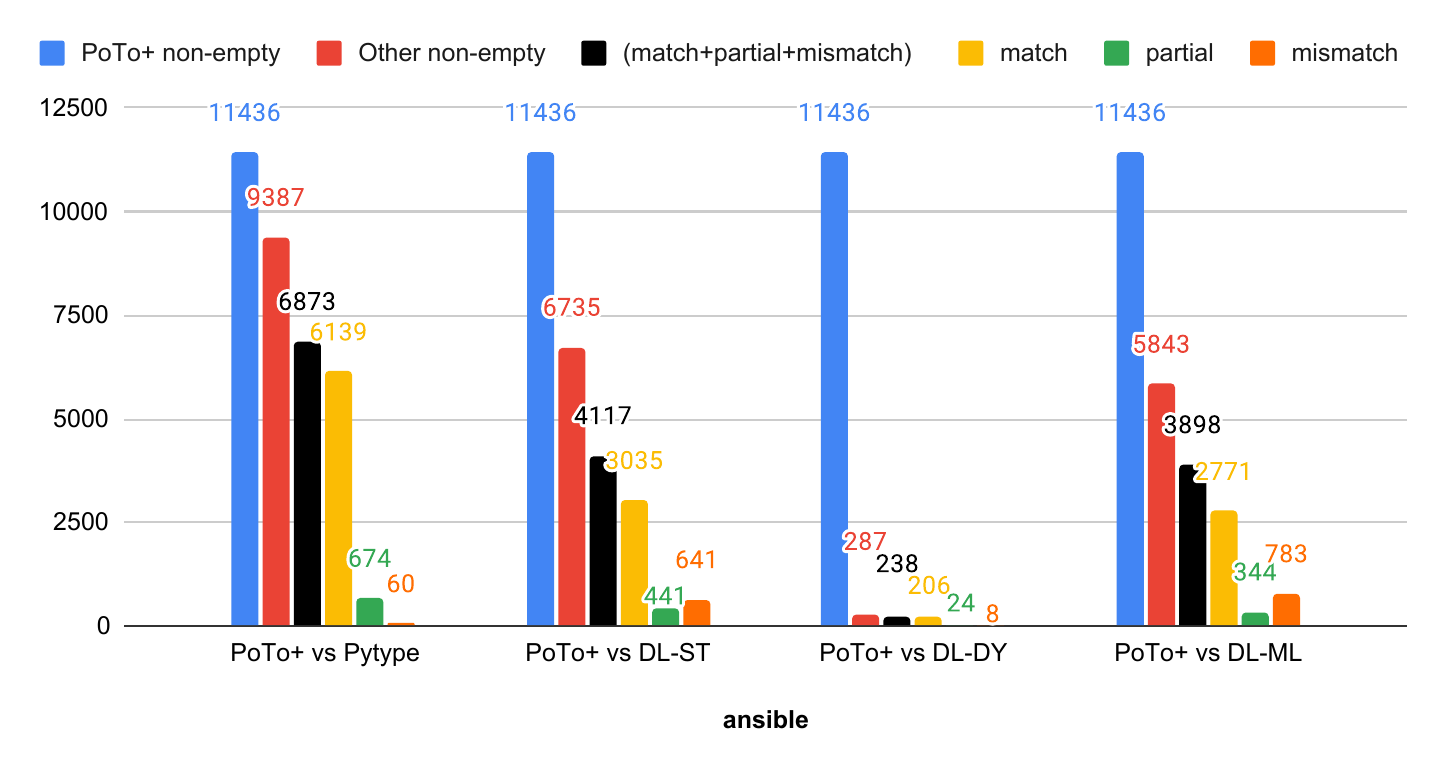}
    \includegraphics[width=0.49\textwidth,trim={15pt 5pt 15pt 20pt},clip]{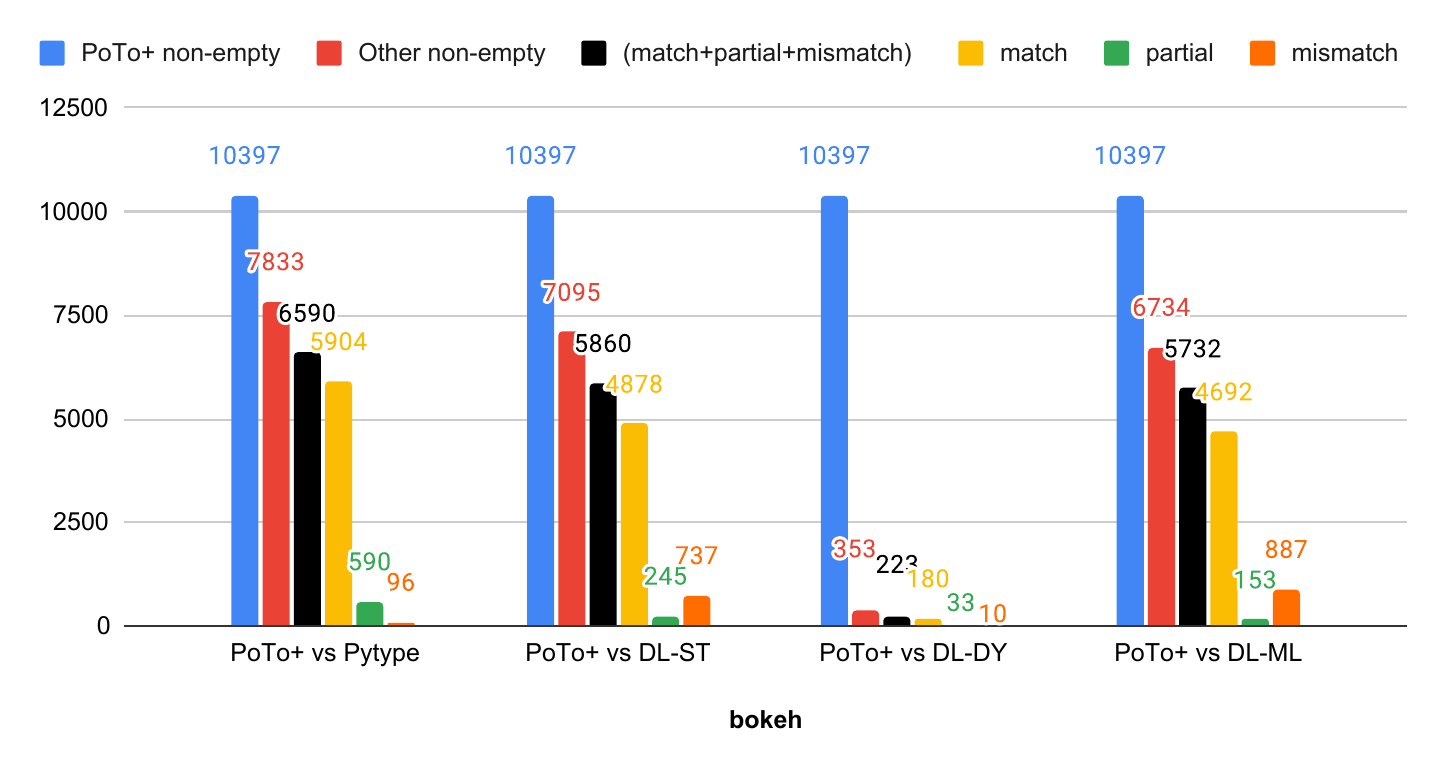}
    \includegraphics[width=0.49\textwidth,trim={15pt 5pt 15pt 20pt},clip]{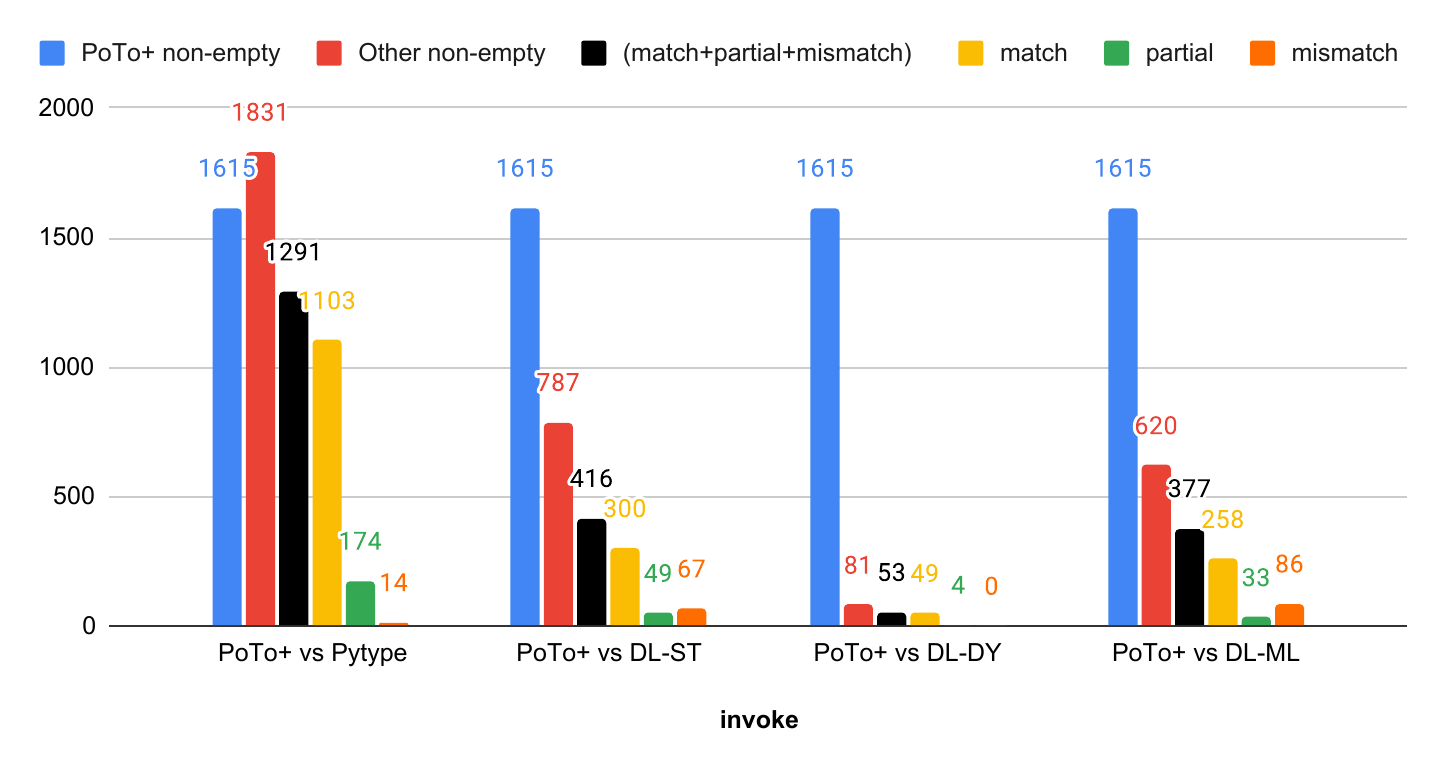}
    \includegraphics[width=0.49\textwidth,trim={15pt 5pt 15pt 20pt},clip]{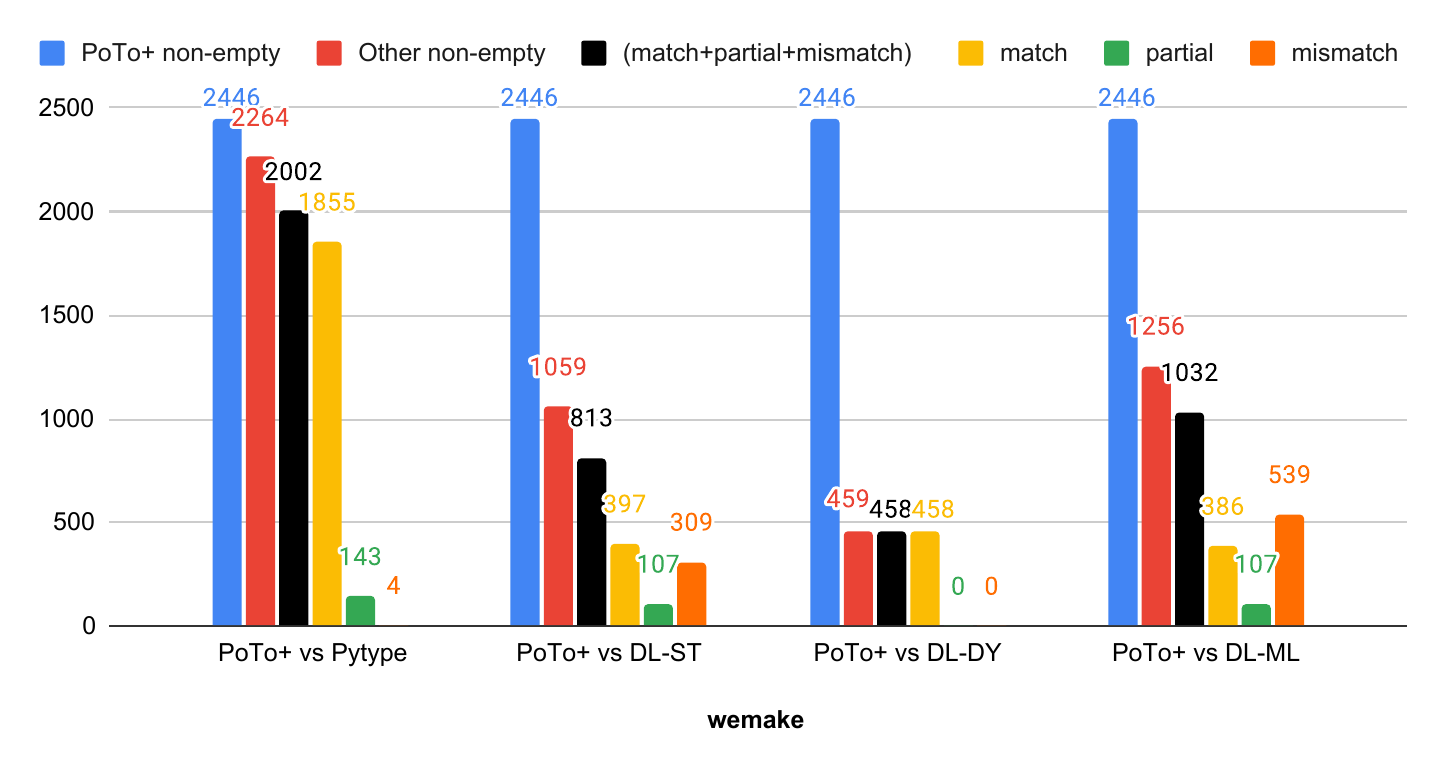}
    \caption{\label{fig:equivalence_average_all}Equivalence comparison between PoTo+ and other type inference techniques (RQ2) for each individual package} 
\end{figure}

\subsection{Difficult-to-Analyze Call Sites}\label{sec:difficult_example}

\begin{lstlisting}
class PublicApi:   
   ...
   def render(self):
     self.setup()
     self.svg.render()

class BaseGraph(PublicApi):
   ...
   def setup(self):
     self.svg = Svg(self)

class Line(BaseGraph):
   ... 
   # Defines several new functions
   # Does not override render or setup
      
# A test:      
line1 = Line()
line1.show_legend = False
line1.fill = True
line1.pretty_print = True
line1.no_prefix = True
line1.x_labels = ['a', 'b', 'c']
line1.add('_', [1, 2, 3])
line1.render()

line2 = Line(show_legend=False, fill=True, pretty_print=True, no_prefix=True, x_labels=['a', 'b', 'c']) 
line2.add('_', [1, 2, 3])
line2.render()

\end{lstlisting}

The listing shows an excerpt of a hierarchy from pygal. The majority of functionality is in base classes \python{PublicApi} and \python{BaseGraph} and the initialization context for graphing objects (e.g., line) consists largely of boolean, string, and numerical values. In this example the two line objects are initialized with the same values, as the test case tests their equivalence, but the important point is the types of encapsulated objects (booleans and lists of strings or integers), not the values themselves. 

The call at line 5 illustrates a performance advantage of PoTo over a 1-object-sensitive analysis. Clearly, PoTo resolves the call to \python{Svg.render()}. A 1-object-sensitive points-to analysis will replicate functions \python{render} and \python{setup} per each one of the line objects hoping to distinguish between the two initialization contexts. However, replication is redundant as field \python{svg} is initialized in each replica of \python{setup} to the same \python{Svg} object, which flows into each replica of \python{render}; the analysis therefore discovers the same call target twice. While this is one example and we do need further studies, we observed similar flows in pygal and other applications --- initialization context is of values of simple types that have no impact on the flow of objects and no impact on the call graph. 


\end{document}